\documentclass[iop]{emulateapj}

\usepackage{natbib, gensymb}
\usepackage{subfigure}
\usepackage{graphicx,amsmath}
\usepackage{amssymb}
\usepackage{color}
\usepackage{natbib}		
\usepackage{url}
\usepackage{multirow}
\usepackage{threeparttable}
\bibpunct{(}{)}{;}{a}{}{,} 
\usepackage{hhline}

\def\Msun{ M_\odot}

\def\Lxuv{ L_{\rm XUV}}

\def\Mp{M_{\rm p}}
\def\Rp{R_{\rm p}}
\def\Sp{S_{\rm p}}

\def\Mearth{ M_\oplus}
\def\Rearth{ R_\oplus}
\def\Searth{ S_\oplus}

\def\p{\mathrm{p}}

\usepackage[normalem]{ulem}
\usepackage{color}
\definecolor{blue}{RGB}{0,0,255}
\definecolor{red}{RGB}{255,0,0}
\definecolor{green}{RGB}{0,200,0}
\definecolor{black}{RGB}{0,0,0}

\def\approxinf{%
  \def\p{%
    \setbox0=\vbox{\hbox{$<$}}%
    \ht0=0.6ex \box0 }%
  \def\s{%
    \vbox{\hbox{$\sim$}}%
  }%
  \mathrel{\raisebox{0.7ex}{%
      \mbox{$\underset{\s}{\p}$}%
    }}%
}

\begin{document}

   \title{Temporal evolution of the high-energy irradiation and water content of TRAPPIST-1 exoplanets.}
   				   
   \author{
   V.~Bourrier\altaffilmark{1},			
   J.~de Wit\altaffilmark{2},			
   E.~Bolmont\altaffilmark{3},			
   V.~Stamenkovi\'c\altaffilmark{4,5},					%
   P.J.~Wheatley\altaffilmark{6},		
   A.J~Burgasser\altaffilmark{7},		
   L.~Delrez\altaffilmark{8},				    
   B.-O.~Demory\altaffilmark{9},					
   D.~Ehrenreich\altaffilmark{1},		
   M.~Gillon\altaffilmark{10}, 			
   E.~Jehin\altaffilmark{10},			
   J.~Leconte\altaffilmark{11},						
   S.M.~Lederer\altaffilmark{12},		
   N.~Lewis\altaffilmark{13}, 						
   A.H.M.J.~Triaud\altaffilmark{14},		
   V.~Van Grootel\altaffilmark{10}					
	}

\shortauthors{V.~Bourrier et al.}
\shorttitle{Irradiation and water loss of TRAPPIST-1 exoplanets}

\altaffiltext{1}{Observatoire de l'Universit\'e de Gen\`eve, 51 chemin des Maillettes, 1290 Sauverny, Switzerland}
\altaffiltext{2}{Department of Earth, Atmospheric and Planetary Sciences, Massachusetts Institute of Technology, 77 Massachusetts Avenue, Cambridge, MA 02139, USA}
\altaffiltext{3}{Laboratoire AIM Paris-Saclay, CEA/DRF - CNRS - Univ. Paris Diderot - IRFU/SAp, Centre de Saclay, F- 91191 Gif-sur-Yvette Cedex, France}
\altaffiltext{4}{Division of Geological and Planetary Sciences, California Institute of Technology, Pasadena, CA 91125 USA}
\altaffiltext{5}{Jet Propulsion Laboratory, California Institute of Technology, Pasadena, CA 91109 USA}
\altaffiltext{6}{Department of Physics, University of Warwick, Coventry CV4 7AL, UK}
\altaffiltext{7}{Center for Astrophysics and Space Science, University of California San Diego, La Jolla, CA 92093, USA}
\altaffiltext{8}{Cavendish Laboratory, J J Thomson Avenue, Cambridge, CB3 0HE, UK}
\altaffiltext{9}{University of Bern, Center for Space and Habitability, Sidlerstrasse 5, CH-3012, Bern, Switzerland}
\altaffiltext{10}{Institut d’Astrophysique et de G\'eophysique, Universit\'e de Li\`ege, All\'ee du 6 Aout 19C, 4000 Li\`ege, Belgium}
\altaffiltext{11}{Laboratoire d’Astrophysique de Bordeaux, Univ. Bordeaux, CNRS, B18N, all\'ee Geoffroy Saint-Hilaire, 33615 Pessac, France}
\altaffiltext{12}{NASA Johnson Space Center, 2101 NASA Parkway, Houston, Texas, 77058, USA}
\altaffiltext{13}{Space Telescope Science Institute, 3700 San Martin Drive, Baltimore, Maryland 21218, USA}
\altaffiltext{14}{Institute of Astronomy, Madingley Road, Cambridge CB3 0HA, UK}

\date{} 
 
\begin{abstract}
The ultracool dwarf star TRAPPIST-1 hosts seven Earth-size transiting planets, some of which could harbour liquid water on their surfaces. UV observations are essential to measure their high-energy irradiation, and to search for photodissociated water escaping from their putative atmospheres. Our new observations of TRAPPIST-1 Ly-$\alpha$ line during the transit of TRAPPIST-1c show an evolution of the star emission over three months, preventing us from assessing the presence of an extended hydrogen exosphere. Based on the current knowledge of the stellar irradiation, we investigated the likely history of water loss in the system. Planets b to d might still be in a runaway phase, and planets within the orbit of TRAPPIST-1g could have lost more than 20 Earth oceans after 8\,Gyr of hydrodynamic escape. However, TRAPPIST-1e to h might have lost less than 3 Earth oceans if hydrodynamic escape stopped once they entered the habitable zone. We caution that these estimates remain limited by the large uncertainty on the planet masses. They likely represent upper limits on the actual water loss because our assumptions maximize the XUV-driven escape, while photodissociation in the upper atmospheres should be the limiting process. Late-stage outgassing could also have contributed significant amounts of water for the outer, more massive planets after they entered the habitable zone. While our results suggest that the outer planets are the best candidates to search for water with the JWST, they also highlight the need for theoretical studies and complementary observations in all wavelength domains to determine the nature of the TRAPPIST-1 planets, and their potential habitability.  
\end{abstract}


\maketitle

\section{Introduction}
\label{intro} 

The TRAPPIST-1 system has been found to host an unprecedented seven Earth-sized planets (\citealt{Gillon2016}, \citealt{Gillon2017}). All seven of the TRAPPIST-1 planets were detected using the transit method \citep{Winn2010}, which allows for the direct determination of their radii (\citealt{Gillon2017}, Table 1). Masses for the TRAPPIST-1 planets (\citealt{Gillon2017}, Table 1) were derived through transit-timing variations (TTV, \citealt{holman2005}). The seventh planet's properties were recently refined by \citet{Luger2017}, who showed that three-body resonances link every planet of this complex system. The combined mass and radius measurements for the TRAPPIST-1 planets are consistent with rocky water-enriched bulk compositions, with TRAPPIST-1 f having a density low enough to harbor up to 50\% of water in its mass. Three of the TRAPPIST-1 planets (e to g) orbit within the habitable zone (HZ) (e.g., \citealt{Kopparapu2013}), where water on a planet's surface is more likely to be in a liquid state. The planets in the TRAPPIST-1 system present unique opportunity thus far for single-system comparative studies aimed at understanding the formation and evolution of terrestrial exoplanet atmospheres.\\

\begin{table*}[htbp]
\begin{center}
\caption{Characteristics of the TRAPPIST-1 exoplanets. Note that the mass of planet h could not be determined with TTV, we therefore computed the mass range for two extreme compositions: 100\% ice and 100\% iron. We chose to base our analysis on the masses derived by \citet{Gillon2017} rather than those of \citet{Wang2017} and \citet{Quarles2017}, because these latter works were still under reviewing at the time of submission of this paper.}
\vspace{0.1cm}
\begin{tabular}{|c|c|c|c|c|c|c|c|}
\hline
Planets 			& b		& c	 	& d 		& e 		& f 		& g 		& h  		\\
\hline
$\Mp (\Mearth)$	& 0.85$\pm$0.72	& 1.38$\pm$0.61	& 0.41$\pm$0.27		& 0.62$\pm$0.58		& 0.68$\pm$0.18	& 1.34$\pm$0.88	& 0.06 -- 0.86 	\\
$\Rp (\Rearth)$		& 1.086	& 1.056	& 0.772		& 0.918		& 1.045	& 1.127	& 0.752 	\\
$\rho_\mathrm{p} (\rho\earth)$ & 0.66$\pm$0.56 & 1.17$\pm$0.53 & 0.89$\pm$0.60 & 0.80$\pm$0.76 & 0.60$\pm$0.17 & 0.94$\pm$0.63 & 0.14 -- 2.02 \\ 
$a_\mathrm{p}$ (au)			& 0.01111	& 0.01521	& 0.02144		& 0.02817		& 0.0371	& 0.0451	& 0.059	\\
\hline
\end{tabular} 
\label{tab_trapp} 
\end{center}
\end{table*}

The atmospheres of terrestrial exoplanets are expected to be diverse and shaped by a number of physical processes (e.g., \citealt{leconte2015}). Observationally probing the TRAPPIST-1 planets over a broad wavelength range from the ultraviolet (UV) to the infrared (IR) provides insights into their current state and the dominant physical processes shaping their atmospheres. Because the TRAPPIST-1 planets transit their host stars as seen from Earth, their atmospheres can be probed via transmission spectroscopy \citep[e.g.][]{seager2000, kaltenegger2009}. The atmospheres of TRAPPIST-1b and c were probed at IR wavelengths by \citet{dewit2016} using the Hubble Space Telescope (HST), which found the atmospheres of these planets to be inconsistent with clear, hydrogen-rich ``primordial" atmospheres. However, a number of plausible scenarios still exist for the atmospheres of TRAPPIST-1b and c, including water-rich and aerosol-laden
atmospheres. A robust interpretation of current and future observations of TRAPPIST-1 at IR wavelengths will require a better understanding of atmospheric chemistry and escape processes shaping these planets, which can be provided by observations at UV wavelengths.  \\

Ultraviolet transit spectroscopy is a powerful way to search for signatures of atmospheric escape from exoplanets. Extended atmospheres of neutral hydrogen have been detected through observations of the stellar Lyman-$\alpha$ line (Ly-$\alpha$, at 1215.67\,\AA) during the transit of Jupiter-mass planets (\citealt{VM2003}; \citealt{Ehrenreich2012}; \citealt{Lecav2010,Lecav2012}) and Neptune-mass planets (\citealt{Kulow2014}, \citealt{Ehrenreich2015}). Because of their spatial extent and kinetic broadening (e.g., \citealt{Ekenback2010}, \citealt{Bourrier_lecav2013}), exospheres transit longer than the lower atmospheric layers probed at optical/infrared wavelengths, and yield deep transit signatures over a large spectral range. The case of the warm Neptune GJ\,436b, in particular, revealed that small planets around cool M dwarfs can support giant exospheres, yielding up to half-eclipses of the star at Ly-$\alpha$ (\citealt{Bourrier2015_GJ436}, \citealt{Bourrier2016}). UV observations of Earth-size planets in a system like TRAPPIST-1 thus offer great perspectives for constraining their atmospheric properties. The faint Ly-$\alpha$ line of this cold M8 star was detected by \citealt{Bourrier2017} (hereafter B17) using the Hubble Space Telescope (HST), with enough light to perform transit spectroscopy. Hints of variations were identified at the time of the transits of inner planets b and c, which could either indicate extended atmospheres of neutral hydrogen or intrinsic stellar variability. The first objective of the present study was to reobserve the TRAPPIST-1 system in the Ly-$\alpha$ line during a TRAPPIST-1c transit, to search for signatures of an extended atmosphere, and to improve our understanding of the high-energy stellar emission.\\

Despite recent efforts (\citealt{France2013}, \citealt{France2016}) our understanding of the atmospheres of exoplanets around M dwarf stars remains limited by the lack of observational and theoretical knowledge about the UV and X-rays spectra of these cool stars. Yet these stars currently offer the best opportunity to detect and characterize Earth-size planets in the habitable zone (HZ). Measuring their UV irradiation is crucial because it impacts the stability and erosion of planetary atmospheres (e.g., \citealt{Lammer2003}, \citealt{Koskinen2007}, \citealt{Bolmont2017}), controls photochemical reactions in the outer atmosphere (\citealt{Miguel2015}), and can further influence the developement and survival of life on a planet surface (see, e.g., \citealt{OmalleyJames2017}, \citealt{Ranjan2017} for TRAPPIST-1). The high present-day X-rays to ultraviolet (XUV) emission from TRAPPIST-1 (\citealt{Wheatley2016}) and the fact that M dwarfs can remain active for several billon years suggest that the atmospheres of the TRAPPIST-1 planets could have been subjected to significant mass loss over the course of their history. Water, in particular, could have been lost through photolysis and atmospheric escape, a process which has been previously studied by \citet{Bolmont2017} for TRAPPIST-1b, c, and d. At the time of their study, only those three planets were known, and the XUV emission of the star was not yet observationally-constrained.\\ 

Our second objective in this paper is to revise the calculations of water-loss for all TRAPPIST-1 planets, benefiting from our improved knowledge of the system architecture (\citealt{Gillon2017}) and of the stellar XUV irradiation (\citealt{Wheatley2016}, B17, and new Ly-$\alpha$ measurements presented in this paper). The planet properties used in our analysis are given in Table~\ref{tab_trapp}. HST observations of TRAPPIST-1 are presented in Sect.~\ref{sec:data_red}, and used in Sect.~\ref{sec:ana_FUV} to study the high-energy stellar emission and its temporal evolution. Sect.~\ref{sec:evol} describes how the stellar irradiation influences the water loss from the planetary atmospheres, while Sect.~\ref{sec:h_loss_crea} addresses the limiting effect of hydrogen production. We discuss the evolution of TRAPPIST-1 exoplanet atmospheres in Sect.~\ref{sec:conclu}.

\section{Observations and data reduction}
\label{sec:data_red} 

\begin{figure}
\centering
\includegraphics[trim=2.cm 3.8cm 9.5cm 3.3cm,clip=true,width=\columnwidth]{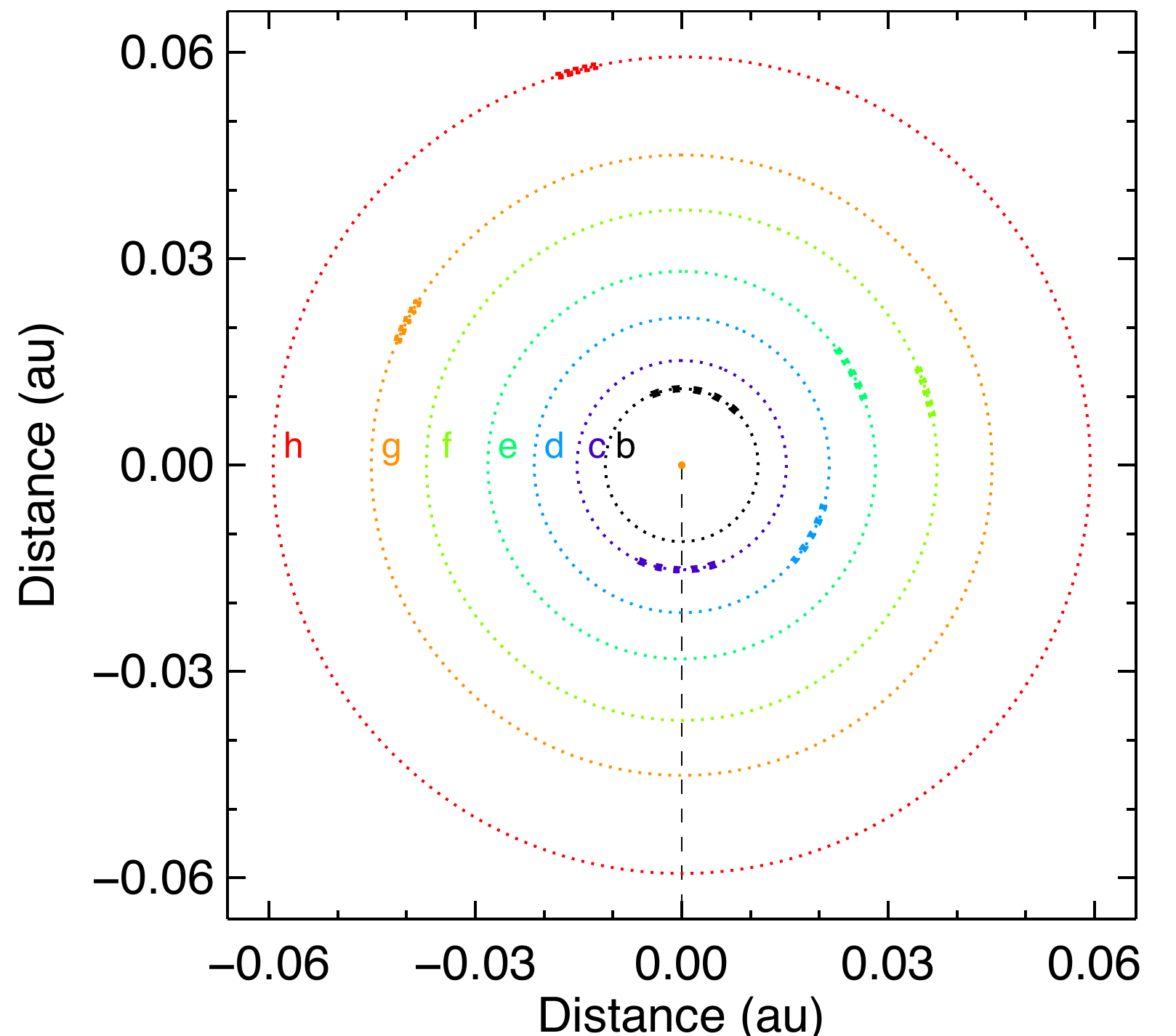}\\
\includegraphics[trim=0cm 2.6cm 1cm 12cm,clip=true,width=\columnwidth]{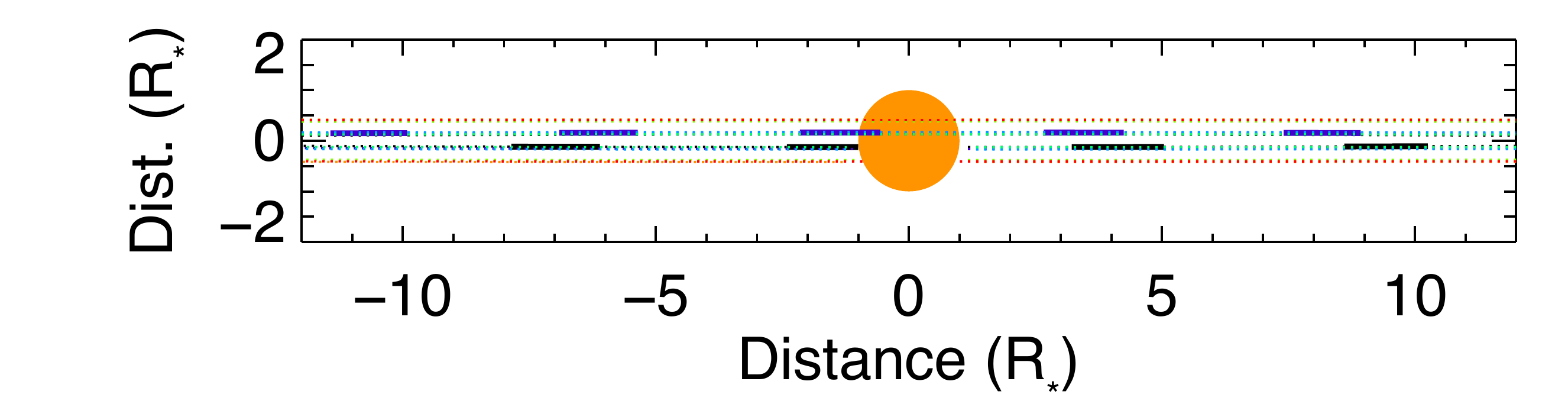}
\caption[]{Orbital positions of the TRAPPIST-1 planets at the time of the HST observations in Visit 4. Each rectangle corresponds to the space covered by a planet during one of the HST orbits. \textit{Upper panel:} View from the above of the planetary system. Planets are moving counterclockwise. The dashed black line indicates the line-of-sight (LOS) toward Earth. Star and orbital trajectory have the correct relative scale. \textit{Lower panel:} View from Earth.}
\label{fig:orb_cov}
\end{figure}

We observed the H\,{\sc i} Ly-$\alpha$ line (1215.6702\,\AA) of TRAPPIST-1 at four independent epochs in 2016, using the Space Telescope Imaging Spectrograph (STIS) instrument on board the HST. The log of these observations is given in Table~\ref{tab:log}. The star was observed at three epochs during Mid-Cycle Program 14493 (PI: V. Bourrier). Two consecutive HST orbits were obtained on 26 September (Visit 1), at a time when none of the seven planets were transiting (all planets were between 135$^{\circ}$ and 330$^{\circ}$ past their last respective transits). A single HST orbit was obtained during the transit of TRAPPIST-1b on 30 September (Visit 2), and another one about 1.7\,h after the transit of TRAPPIST-1c on 23 November (Visit 3). No other planets were close to transiting during Visits 2 and 3. Results of this reconnaissance program were published in B17. We obtained five new orbits on December 25, 2016 during the GO/DD Program 14900 (PI: J.~de Wit). Visit 4 was scheduled to include a TRAPPIST-1c transit, to search for the signature of a hydrogen exosphere around the planet. The configuration of the planetary system at the time of Visit 4 is shown in Fig.~\ref{fig:orb_cov}. Note that because of occultations by the Earth and the time required to acquire the target star with the HST, about a third of an HST orbit could be spent observing TRAPPIST-1 at Lyman-$\alpha$. \\

\begin{table}[tbh]
\centering
\caption{Log of TRAPPIST-1 Ly-$\alpha$ observations in 2016.}
\begin{tabular}{lccc}
\hline
\hline
\noalign{\smallskip}
Visit & Date & \multicolumn{2}{c}{Time (UT)}   \\
      &      & Start & End                                \\
\noalign{\smallskip}
\hline
1 & Sept-26     & 02:51:50   & 04:59:45 \\
2 & Sept-30     & 22:55:05   &  23:27:38\\
3 & Nov-23      & 20:56:10   &  21:28:35\\
4 & Dec-25      &  03:17:58  &  10:10:17\\
\noalign{\smallskip}
\hline
\hline
\end{tabular}
\label{tab:log}
\end{table}

\begin{figure}     
\includegraphics[trim=1.cm 3cm 6cm 6cm,clip=true,width=\columnwidth]{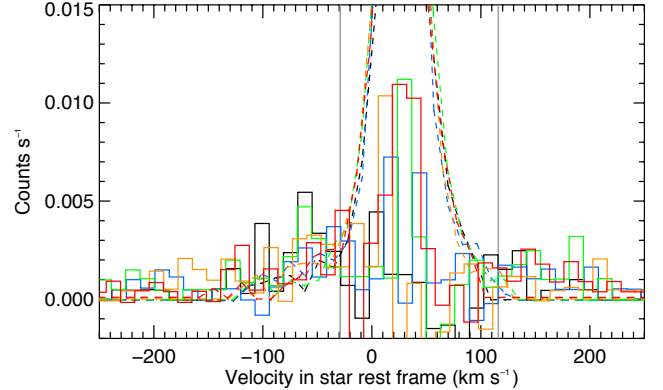}
\caption[]{Raw spectra of the stellar Ly-$\alpha$ line in Visit 4 (solid line histogram), after correction from the geocoronal emission line (superimposed as a dashed line). Colors correspond to HST orbits at consecutive orbital phases (increasing from black, blue, green, orange, red). Gray vertical lines indicate the range excluded from our analysis, where the airglow is so strong that its correction results in spurious flux values.}
\label{fig:airglow}
\end{figure}

All four visits made use of STIS Far Ultraviolet Multi-Anode Microchannel Array (FUV-MAMA) detector, and the G140M grating at 1222\,\AA. Data were reduced with the CALSTIS pipeline. In the region of the Ly-$\alpha$ line the background is dominated by the Earth's geocoronal airglow emission (\citealt{VM2003}). The error bars in the final 1D spectra account for the uncertainty in the airglow flux, but the correction performed by the pipeline can yield spurious flux values where the airglow is much stronger than the stellar line. The position, amplitude, and width of the airglow line profile varies in strength and position with the epoch of observation (e.g. \citealt{Bourrier2016_HD976}), and we thus excluded from our analysis the contaminated ranges [-3 ; 129]\,km\,s$^{-1}$ (Visit 1), [-4 ; 102]\,km\,s$^{-1}$ (Visit 2), [-8 ; 110]\,km\,s$^{-1}$ (Visit 3), and [-29 ; 116]\,km\,s$^{-1}$ (Visit 4), defined in the star rest frame. Airglow is much stronger in Visit 4 because TRAPPIST-1 was nearly four months past opposition (see Fig.~\ref{fig:airglow}), and we found that the stellar spectrum between -150 and -29\,km\,s$^{-1}$ depended on the areas of the 2D images used to build the background profile. The background is extracted and averaged automatically by the pipeline from two regions above and below the spectrum. For the FUV-MAMA D1 aperture used in Visit 4, the standard regions are 5 pixels wide and located $\pm$30 pixels from the spectrum along the cross-dispersion axis. We varied this distance and found that individual airglow-corrected exposures showed differences in the blue wing of the stellar Ly-$\alpha$ line when the background was extracted from regions farther than $\sim$25 pixels from the spectrum. We thus limited the effect of airglow contamination by measuring an accurate local background within extraction regions that extend between 6 and 20 pixels from each side of the spectrum.\\

\begin{figure}     
\includegraphics[trim=2cm 6.5cm 1cm 10cm,clip=true,width=\columnwidth]{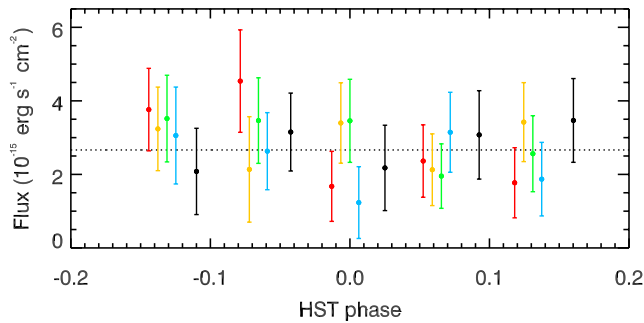}
\caption[]{Ly-$\alpha$ fluxes for all sub-exposures in Visit 4, integrated over the entire line (excluding the airglow range) and phase-folded on the HST orbital period. No variations caused by the telescope breathing were detected, with no significant deviations from the mean flux (black dashed line). Colors correspond to HST orbits at consecutive orbital phases (increasing from black, blue, green, orange, red).}
\label{fig:breath_flux}
\end{figure}

Data in each orbit, obtained in time-tagged mode, were divided into five shorter sub-exposures (varying from 380 to 450\,s depending on the duration of the initial exposure). This allowed us to check for variations at short time scales within a given HST orbit caused by the telescope “breathing” (e.g., \citealt{Bourrier2013}). We modeled the breathing effect using either a Fourier series decomposition (\citealt{Bourrier2016_HD976}) or a polynomial function (e.g., \citealt{Ehrenreich2012}). The model was fitted to the flux integrated over the entire Ly-$\alpha$ line, excluding the range contaminated by airglow emission, and using the Bayesian Information Criterion (BIC) as a merit function (\citealt{Crossfield2012}). No variations caused by the telescope breathing were detected in any of the visits (see Fig.~\ref{fig:breath_flux} for Visit 4), most likely because it is dominated by the photon noise from the very faint Ly-$\alpha$ line and because there are not enough HST orbits to sample its variations properly in Visits 1 to 3. \\

\section{Analysis of TRAPPIST-1 FUV observations}
\label{sec:ana_FUV}

\subsection{Long-term evolution of the stellar Ly-$\alpha$ line}
\label{sec:long_term}

\begin{figure}     
\includegraphics[trim=1.8cm 7.5cm 10cm 6.4cm,clip=true,width=\columnwidth]{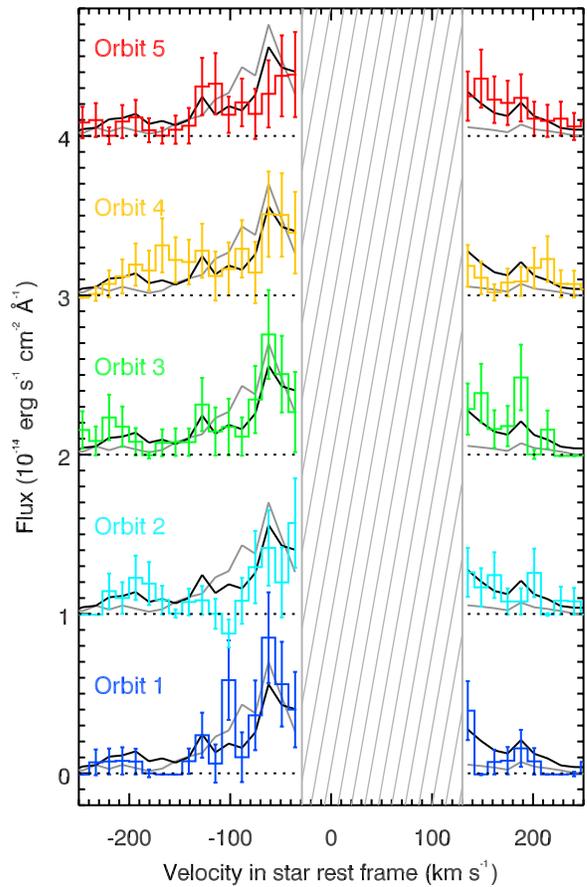}
\caption[]{Ly-$\alpha$ line spectra in Visit 4, overplotted with their average over the visit (black spectrum), and with the reference spectrum from B17 (grey spectrum). Spectra were shifted along the vertical axis (dotted black lines indicate the null level in each orbit). The dashed range is contaminated by airglow emission. TRAPPIST-1 Ly-$\alpha$ line is so faint that in some pixels no photons were detected over the duration of the exposure.}
\label{fig:Spectres_V4_V123}
\end{figure}

A high-quality reference spectrum for the intrinsic stellar Ly-$\alpha$ line of TRAPPIST-1 was built by B17 as the average of all spectra obtained in Visits 1, 2, and 3, excluding the spectral ranges contaminated by airglow emission or showing hints of flux variations. We first compared the Ly-$\alpha$ line spectra obtained in each exposure of Visit 4 with this reference spectrum, searching for absolute flux variations over ranges covering more than STIS spectral resolution (about two pixels). As can be seen in Fig.\ref{fig:Spectres_V4_V123} the flux in the red wing of Visit 4 spectra is systematically higher than or equal to the reference spectrum, with a significant ($>$3$\sigma$) increase in orbits 2, 3, and 5. Similarly the reference spectrum shows very little emission in the blue wing at velocities lower than about -160\,km\,s$^{-1}$, whereas the flux in Visit 4 spectra is systematically higher than or equal to the reference in this range, with a marginal ($>$2$\sigma$) increase in orbits 2, 3, and 4. Therefore it comes as a surprise that Visit 4 displays an overall lower flux in the blue wing between about -120 and -55\,km\,s$^{-1}$, with significant decreases in orbits 2 and 5 compared to the reference spectrum. \\

\begin{figure}     
\includegraphics[trim=1.9cm 4.5cm 3.5cm 10.4cm,clip=true,width=\columnwidth]{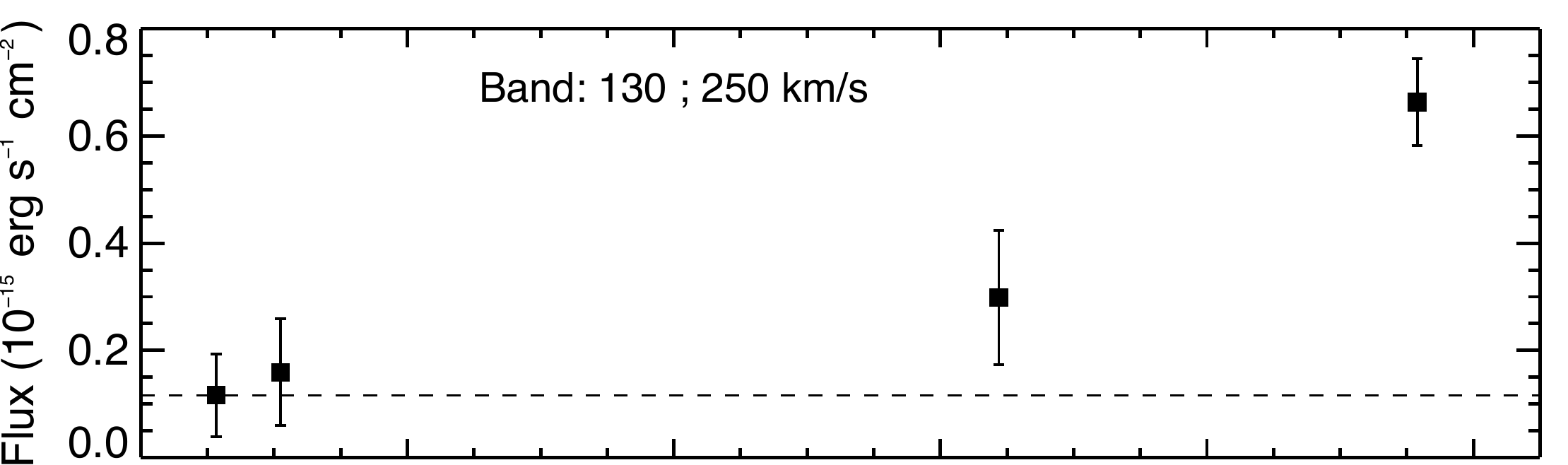}
\includegraphics[trim=1.9cm 4.5cm 3.5cm 10.4cm,clip=true,width=\columnwidth]{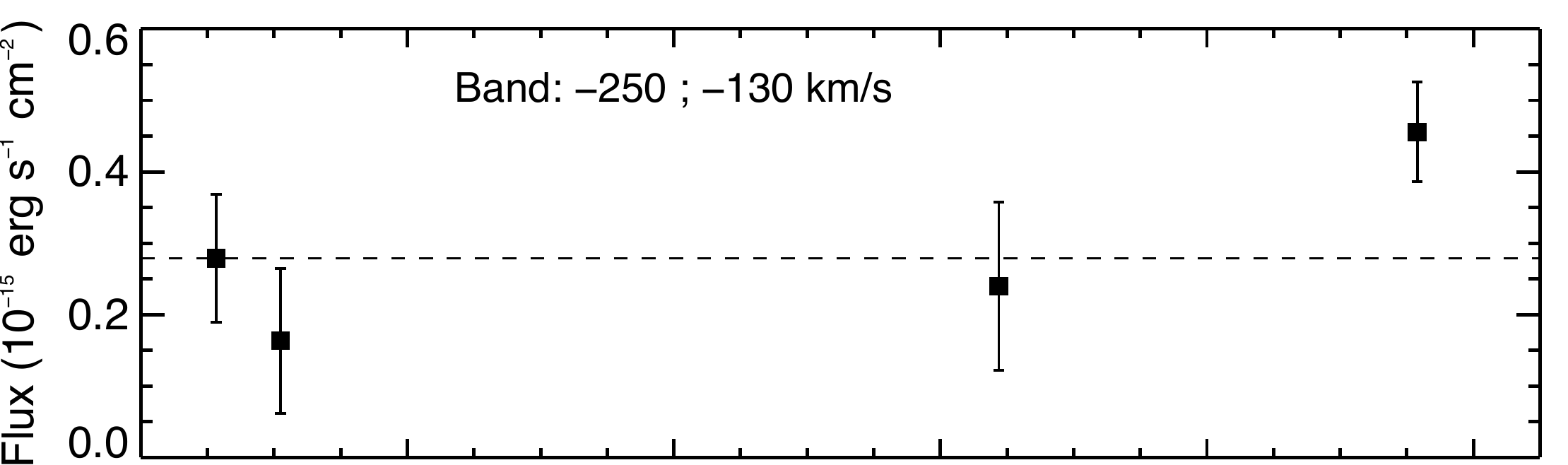}
\includegraphics[trim=1.9cm 4.5cm 3.5cm 10.4cm,clip=true,width=\columnwidth]{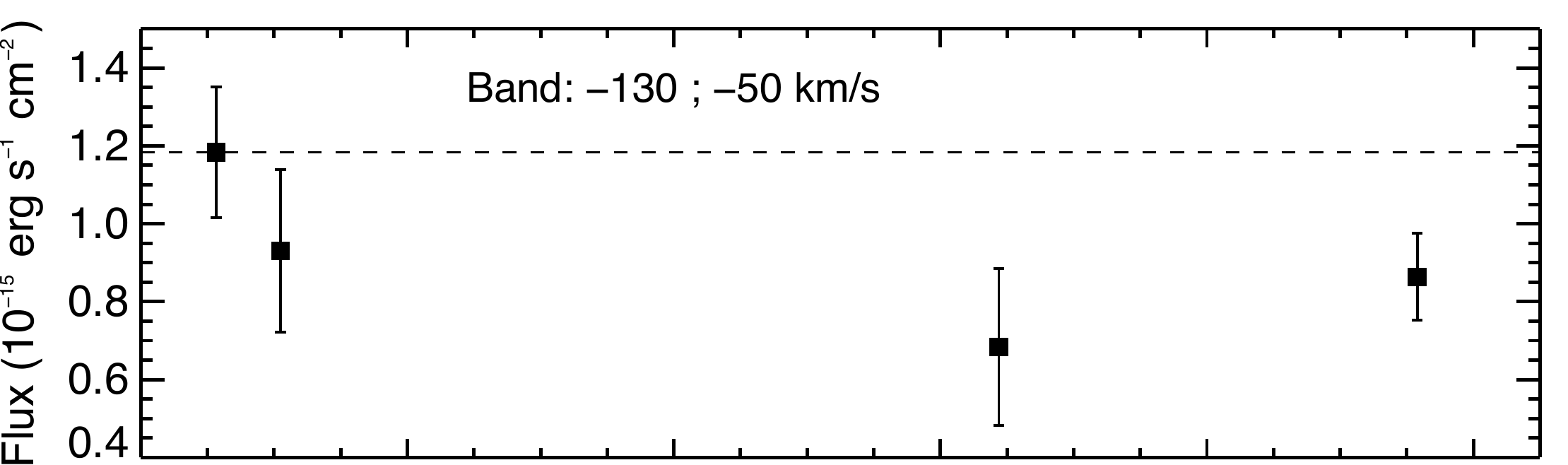}
\includegraphics[trim=1.9cm 3cm 3.5cm 10.4cm,clip=true,width=\columnwidth]{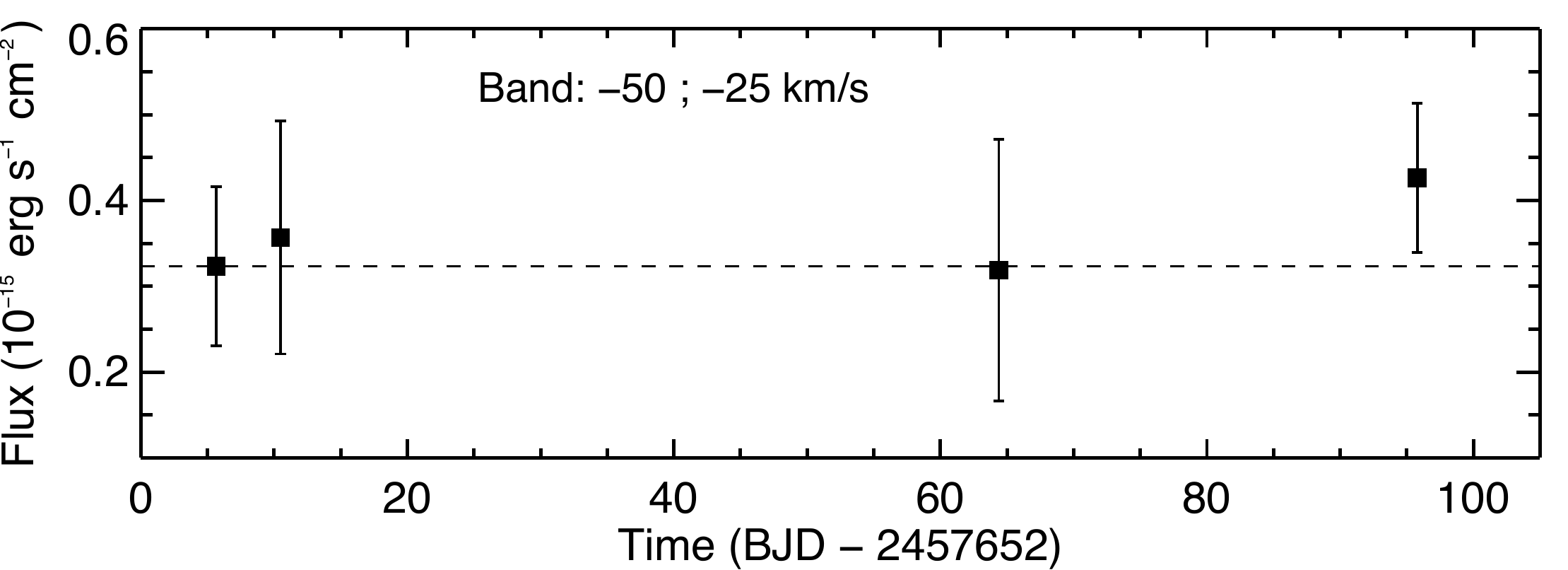}
\caption[]{Evolution of the Ly-$\alpha$ flux over time, integrated in four complementary bands (indicated in each panel). All spectra have been interpolated over a common wavelength table before being averaged in each visit. Each point thus corresponds to the mean spectrum over a visit. The dashed line indicates the flux level in Visit 1, measured outside of any planet transit.}
\label{fig:LC_BJD}
\end{figure}

To investigate the source of these differences we studied the evolution of the Ly-$\alpha$ line over the three months span of our observations, averaging all spectra within each visit, and integrating them in four complementary spectral bands (Fig.~\ref{fig:LC_BJD}). In agreement with the above spectral analysis, the flux in the symmetric wing bands ($\pm$[130 ; 250]\,km\,s$^{-1}$) did not vary significantly from Visit 1 to Visit 3 (top panels in Fig.~\ref{fig:LC_BJD}) but increased noticeably in Visit 4. This variation likely traces an increase in the emission of the intrinsic stellar Ly-$\alpha$ line (Sect.~\ref{sec:cor_struc}). However, while we do not expect similar absolute flux levels in the observed wing bands because of interstellar medium (ISM) absorption in the red wing of the Ly-$\alpha$ line (see B17), it is surprising that the relative flux increase in Visit 4 is much larger and more significant in the red wing than in the blue wing. The flux in the blue wing even shows a marginal decline at lower velocities ([-130 ; -50]\,km\,s$^{-1}$) from Visit 1 to Visit 4 (third panel in Fig.~\ref{fig:LC_BJD}), while the flux at velocities closer to the Ly-$\alpha$ line core ([-50 ; -25]\,km\,s$^{-1}$) remained at about the same flux level (bottom panel in Fig.~\ref{fig:LC_BJD}). \\

This comparison suggests that the shape of the line evolved between Visit 1 and Visit 4, with a change in the spectral balance of the Ly-$\alpha$ line flux between the blue and the red wings. The search for absorption signatures possibly caused by the transit of TRAPPIST-1c in Visit 4 is made difficult by this evolution, since the spectra from Visits 1-3 cannot be used as a reference for the out-of-transit stellar line. We investigate this question in more details in Sect.~\ref{sec:short_term}.\\

\subsection{High-energy stellar emission}
\label{sec:cor_struc}

To further study the evolution of the stellar Ly-$\alpha$ line, we sought to reconstruct its intrinsic profile at the time of Visit 4 using the same approach as in B17. We assumed a Gaussian line profile, which was absorbed by the ISM using the column density derived in B17. The model was then convolved by the STIS line spread function (LSF), and compared with the average of all spectra in Visit 4. We excluded the pixels beyond $\pm$250\,km\,s$^{-1}$ from the fit, where the wings of the line become too faint. We also excluded the core of the line fully absorbed by the ISM and possibly biased by the airglow correction. \\

\begin{figure}     
\includegraphics[trim=0.1cm 5.5cm 1.7cm 10.5cm,clip=true,width=\columnwidth]{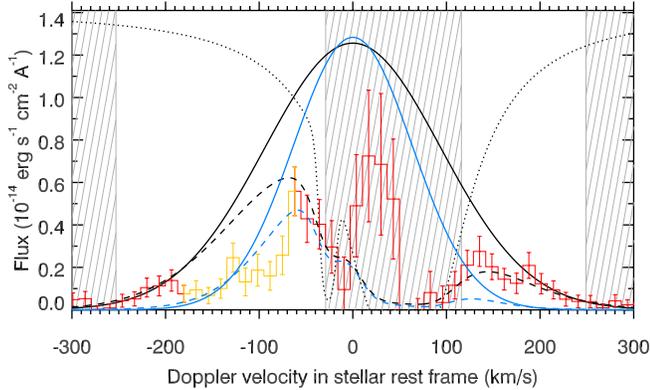}
\caption[]{Ly-$\alpha$ line profiles of TRAPPIST-1. Solid-line profiles correspond to our best estimates for the theoretical intrinsic Ly-$\alpha$ line in Visit 1-3 (blue) and in Visit 4 (black). They yield the dashed-line profiles after ISM absorption and convolution by STIS LSF. ISM absorption profile in the range 0-1 has been scaled to the vertical axis range and plotted as a dotted black line. The dashed-line profile in Visit 4 was fitted to the observations (red histogram, equal to the average of all spectra in Visit 4) outside of the hatched regions, and excluding the variable range between -187 and -55\,km\,s$^{-1}$ (highlighted in orange). Note that the model fits the observations well, even in the range contaminated by the airglow (except where it is strongest between 0 and 50\,km\,s$^{-1}$).}
\label{fig:theo_spec}
\end{figure}

Contrary to the reconstruction performed in B17 on Visit 1-3 reference spectrum, in Visit 4, we found that it was not possible to fit the entire line profile well with our theoretical Gaussian model. Trying out different spectral ranges for the reconstruction, we found that a good fit was obtained when excluding the band [-190 ; -55]\,km\,s$^{-1}$, even in the range contaminated by the airglow (Fig.~\ref{fig:theo_spec}). This best-fit line profile for Visit 4 peaks at about the same flux level as the reference spectrum in Visit 1-3 but displays much broader wings, which is consistent with the stability of the Ly-$\alpha$ line core and the variability of its wings noted in Sect.~\ref{sec:long_term}. Assuming this best-fit model is correct, it would suggest that both wings of the Ly-$\alpha$ line have increased similarly from Visit 1-3 to Visit 4, but that the band [-190 ; -55]\,km\,s$^{-1}$ is re-absorbed by an unknown source in this epoch. Alternatively, the intrinsic Ly-$\alpha$ line of TRAPPIST-1 could have become asymmetric in Visit 4. Both scenarios would explain why the line profile observed in Visit 4 appears unbalanced between the blue and the red wing (see Sect.~\ref{sec:long_term} and Fig.~\ref{fig:LC_BJD}). We further investigate this question in Sect.~\ref{sec:short_term}.

The Lyman-$\alpha$ line arises from different regions of the stellar atmosphere, ranging from the low-flux wings of the line formed in the colder regions of the lower chromosphere, to the core of the line, which is emitted by the hot transition region between the upper chromosphere of the star and its corona. M dwarfs display a lower chromospheric emission than earlier-type stars, but equivalent amounts of emission from the transition region and the corona (see \citealt{Youngblood2016} and references). This trend might be even more pronounced for late-type M dwarfs like TRAPPIST-1, since B17 suggested that this ultracool dwarf might have a weak chromosphere compared to its transition region and corona, based on its Lyman-$\alpha$ and X-ray emission (\citealt{Wheatley2016}) and the shape of its Ly-$\alpha$ line. Interestingly though, the broader wings of TRAPPIST-1 Ly-$\alpha$ line in Visit 4 might trace an increase in the temperature and emission of the stellar chromosphere. \\

\begin{figure}     
\includegraphics[trim=1.9cm 6.9cm 2.cm 15.1cm,clip=true,width=\columnwidth]{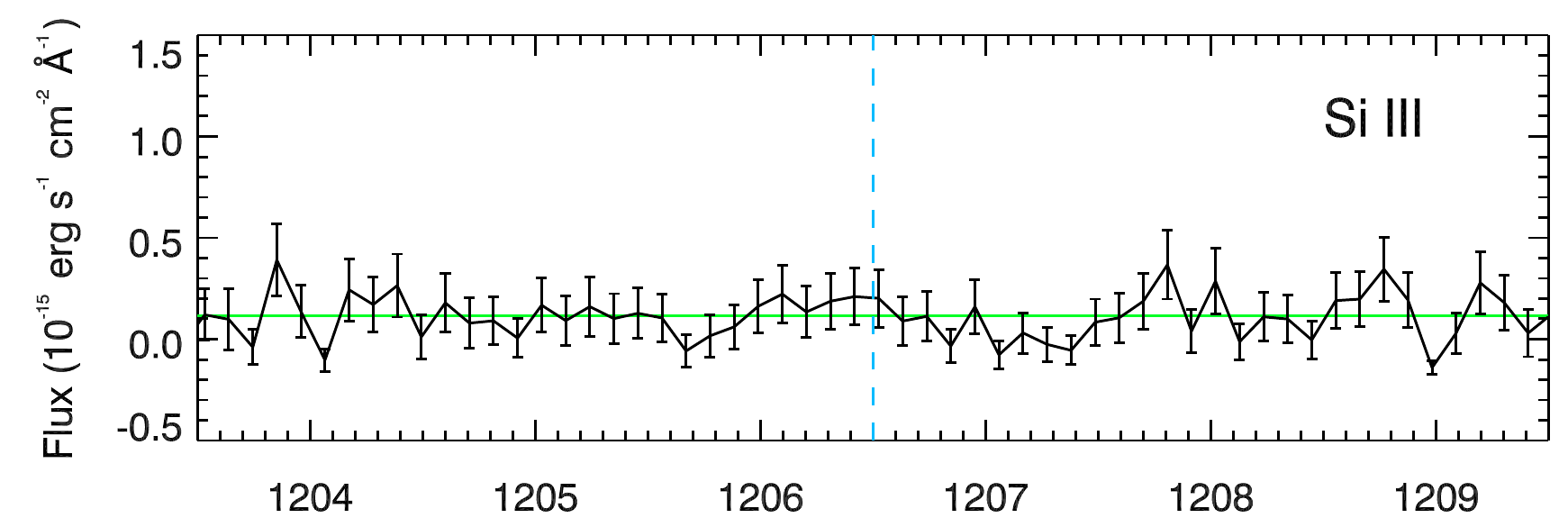}
\includegraphics[trim=1.9cm 6.cm 2cm 15.1cm,clip=true,width=\columnwidth]{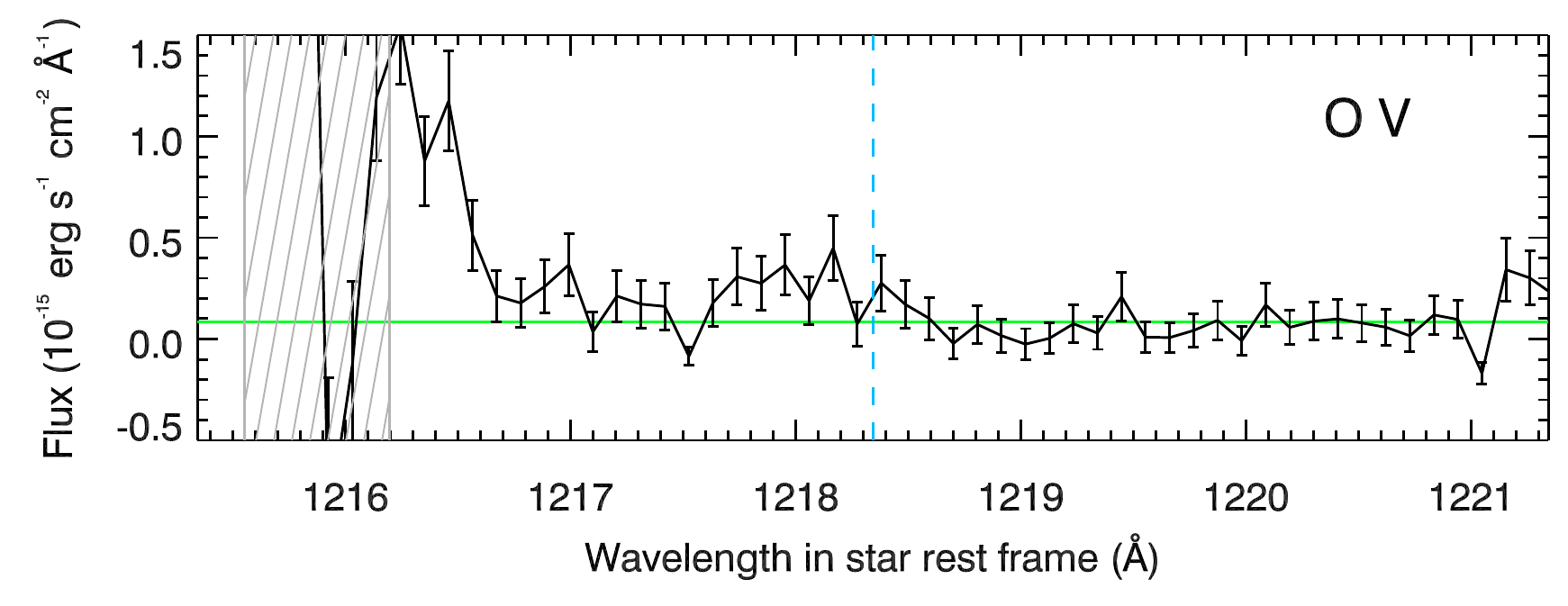}
\includegraphics[trim=1.9cm 6cm 2cm 15.1cm,clip=true,width=\columnwidth]{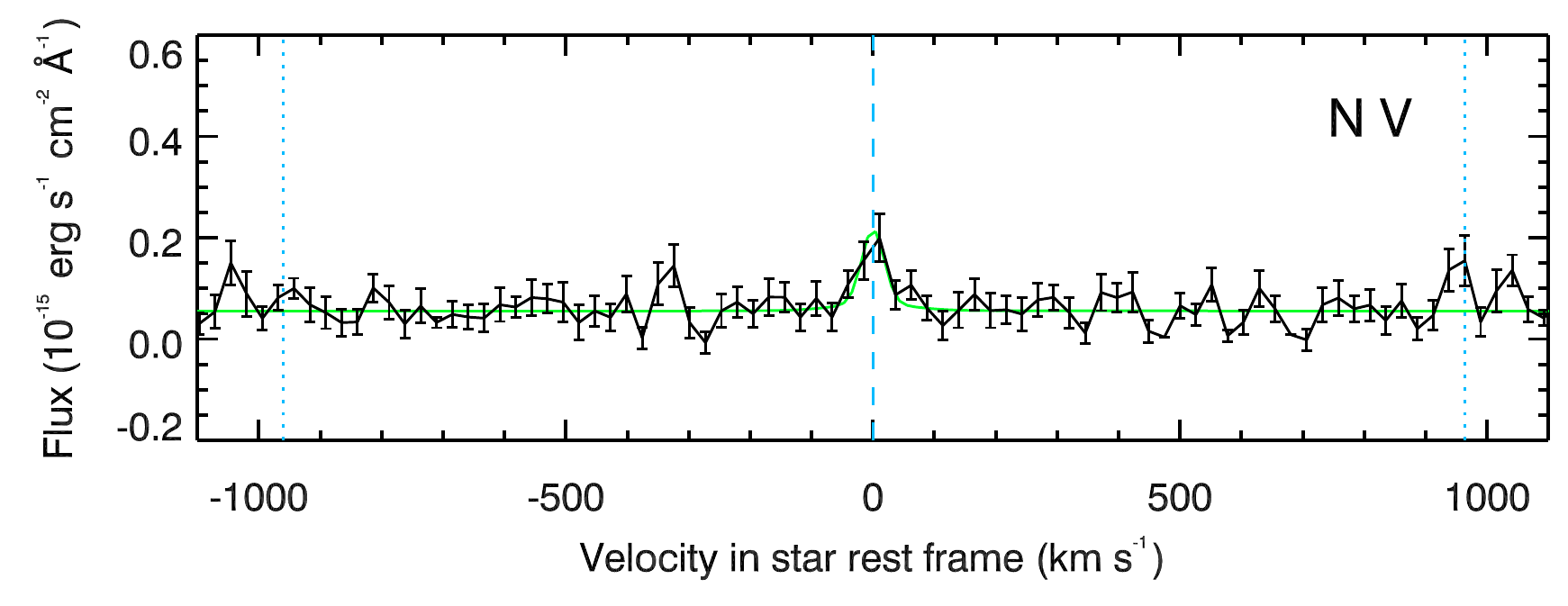}
\caption[]{Average spectrum of TRAPPPIST-1 in the star rest frame in the ranges of the Si\,{\sc iii} line (top panel), the O\,{\sc v} line (middle panel), and the N\,{\sc v} doublet (bottom panel). Pixels are binned by two for the sake of clarity. The blue dashed lines indicate the rest wavelength of the stellar lines. The green line is the mean flux in the range (excluding the region blueward of the O\,{\sc v} line, since it is dominated by the red wing of the Ly-$\alpha$ line). In the bottom panel, the lines of the N\,{\sc v} doublet have been averaged in velocity space and fitted with a Gaussian model. Each N\,{\sc v} line is indicated by a dotted blue line at the velocity of its transition relative to the other line of the doublet.}
\label{fig:stellar_lines}
\end{figure}

In addition to the Ly-$\alpha$ line, the spectral range of the STIS/G140M grating covers the Si\,{\sc iii} (1206.5\,\AA) and O\,{\sc v} (1218.3\,\AA) transitions, and the N\,{\sc v} doublet (1242.8\,\AA\ and 1238.8\,\AA). We averaged our nine STIS spectra of TRAPPIST-1 to search for these stellar emission lines, and detected the N\,{\sc v} doublet (Fig.~\ref{fig:stellar_lines}). The two lines of the doublet were averaged and fitted with a Gaussian profile. Assuming that the width of the line is controlled by thermal broadening, we obtained a best-fit temperature on the order of 3$\times$10$^{5}$\,K, which is close to the peak emissivity of the N\,{\sc v} lines at 2$\times$10$^{5}$\,K. We further derived a total flux in the doublet of about 7.3$\times$10$^{-17}$\,erg\,s$^{-1}$\,cm$^{-2}$. This line strength is consistent with the fits to the X-ray spectrum of TRAPPIST-1 by \citealt{Wheatley2016}, which predict N\,{\sc v} line strengths of 1$\times$10$^{-17}$ and 100$\times$10$^{-17}$\,erg\,s$^{-1}$\,cm$^{-2}$ for two different models designed to span the possible range of EUV luminosities (the APEC and {\it cemekl} models respectively). The dispersion is larger in the regions of the other stellar lines and there is no clear evidence for their detection. More observations will be required to characterize the chromospheric and coronal emission from TRAPPIST-1. \\

\subsection{The mysterious shape of TRAPPIST-1 Ly-$\alpha$ line}
\label{sec:short_term}

\begin{figure}     
\includegraphics[trim=1.9cm 4.5cm 2.cm 10.4cm,clip=true,width=\columnwidth]{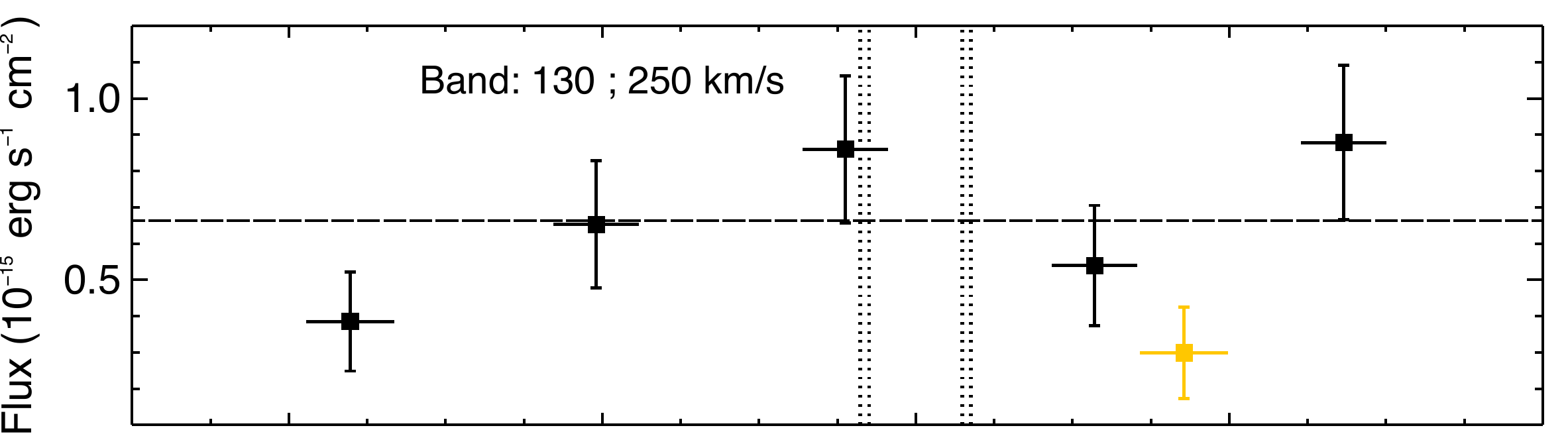}
\includegraphics[trim=1.9cm 4.5cm 2cm 10.4cm,clip=true,width=\columnwidth]{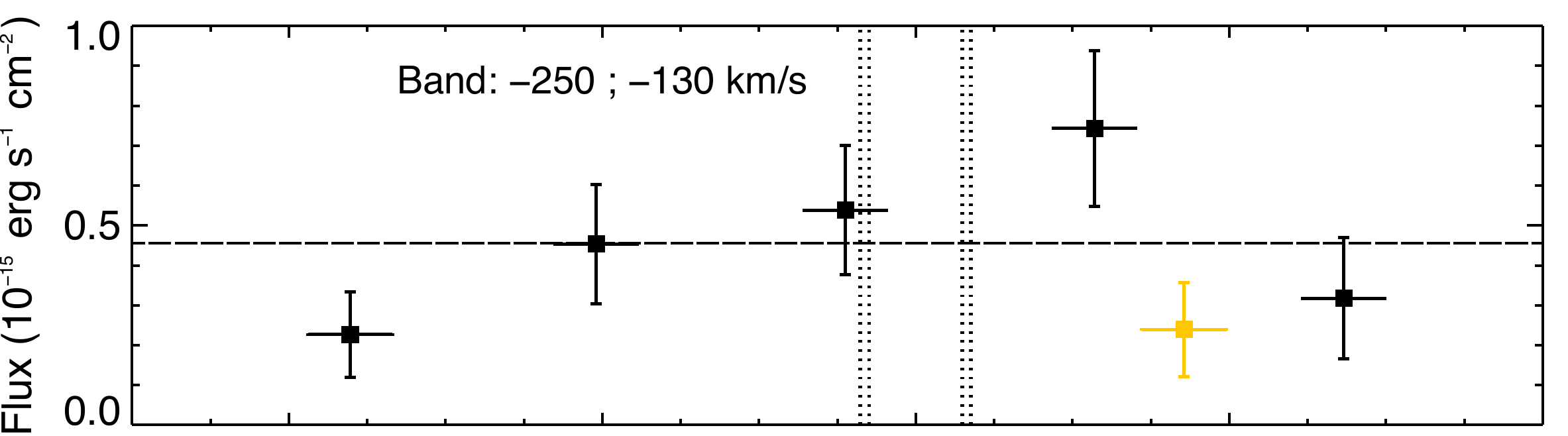}
\includegraphics[trim=1.9cm 4.5cm 2cm 10.4cm,clip=true,width=\columnwidth]{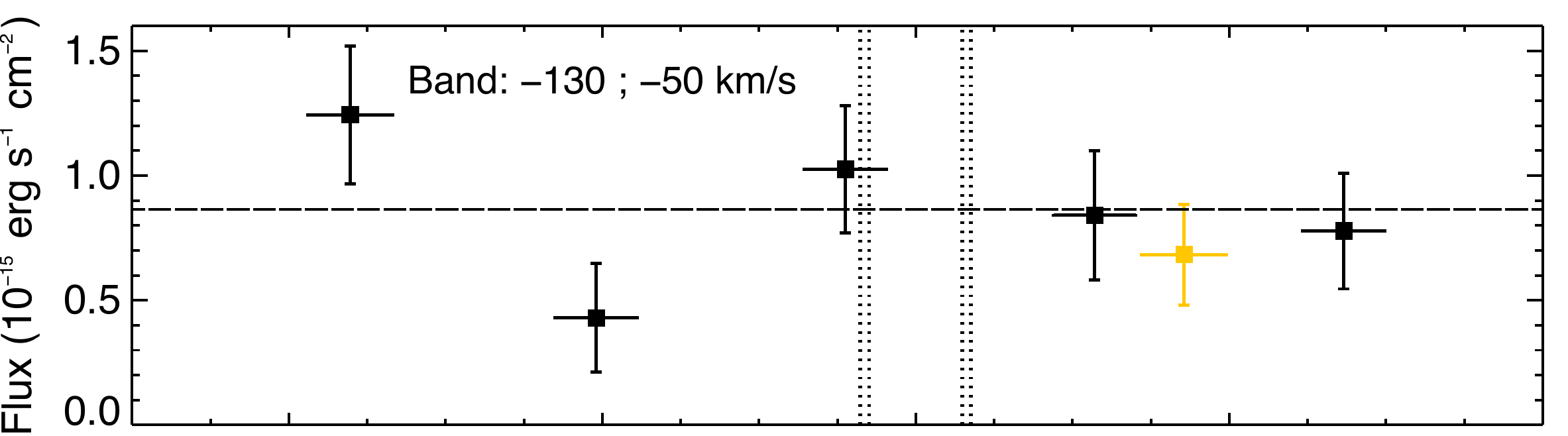}
\includegraphics[trim=1.9cm 3cm 2cm 10.4cm,clip=true,width=\columnwidth]{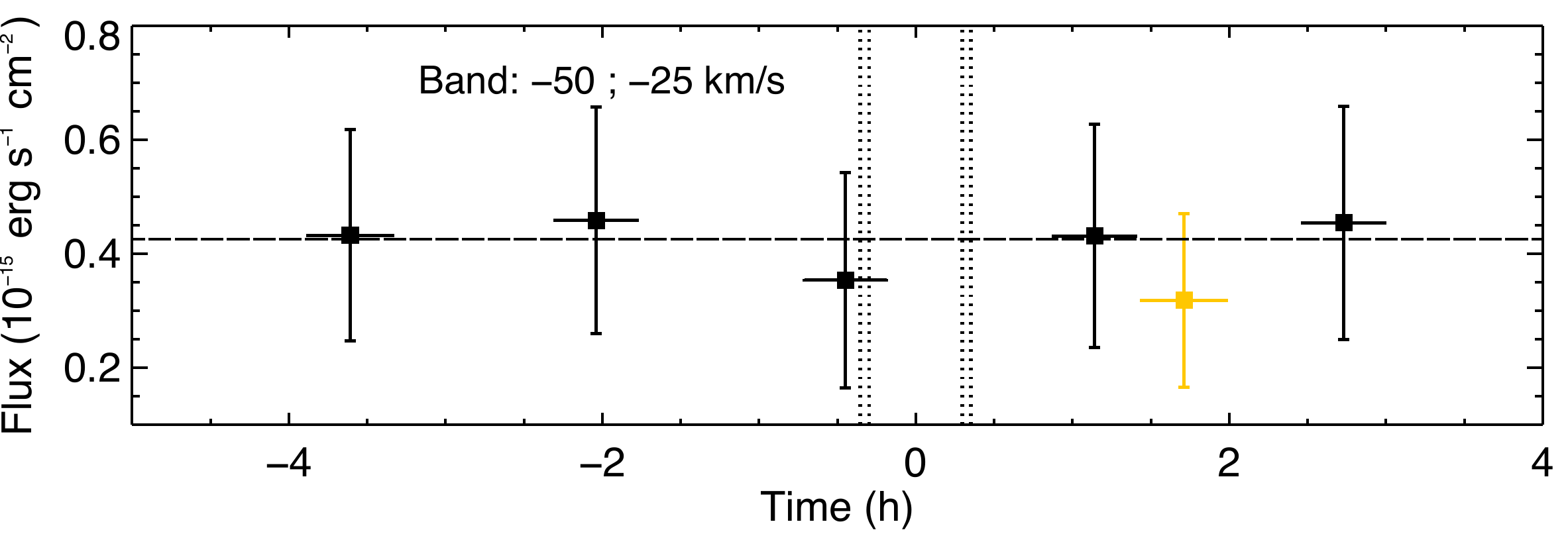}
\caption[]{Ly-$\alpha$ flux integrated in the same complementary bands as in Fig.~\ref{fig:LC_BJD}, and plotted as a function of time relative to the transit of TRAPPIST-1c. Vertical dotted lines indicate the transit contact times. Black points correspond to Visit 4, and the orange point for Visit 3. The dashed line is the mean flux in Visit 4.}
\label{fig:LC_trC}
\end{figure}

B17 reported a hint of absorption in the blue wing of the Ly-$\alpha$ line in Visit 3, possibly caused by a hydrogen exosphere trailing TRAPPIST-1c. The long-term evolution of the intrinsic Ly-$\alpha$ line prevents us from comparing the different visits from one another, and we thus took the average of all spectra in Visit 4 as a reference to search for the presence of an exosphere in this epoch. We highlight the long-term variability in Fig.~\ref{fig:LC_trC}, where we plotted the flux integrated in the same spectral bands as in Fig.~\ref{fig:LC_BJD} but phase-folded over the TRAPPIST-1c orbital period. The spectrum measured in Visit 3 shows a lower flux level at high velocities in the wing bands compared to the average flux in Visit 4 but similar flux levels in other parts of the line, consistent with the changes in the intrinsic line shape discussed in Sect.~\ref{sec:long_term}. No significant deviations to the average spectrum were found in Visit 4, and in particular no variations that would be consistent with the transit of an extended exosphere surrounding TRAPPIST-1c. More observations will be required to assess the possible short-term variability in the Ly-$\alpha$ line and to search for residual absorption signatures.\\

If the Ly-$\alpha$ line model derived in Sect.~\ref{sec:cor_struc} corresponds to the actual intrinsic line of the star in Visit 4, what is the origin of the much lower flux observed in the band [-190 ; -70]\,km\,s$^{-1}$ (Fig.~\ref{fig:theo_spec}) ? Despite our careful extraction of the stellar spectrum (Sect.~\ref{sec:data_red}), it is possible that the stronger airglow in this epoch was overcorrected at some wavelengths. However, the airglow becomes negligible beyond about -100\,km\,s$^{-1}$, and therefore cannot explain the lower flux observed at larger velocities. If confirmed, this feature might imply that colder hydrogen gas is moving away from the star at high velocities and is absorbing about half of the Ly-$\alpha$ flux in this velocity range. This absorption is unlikely to originate from TRAPPIST-1c alone, as it occurs in all orbits of Visit 4 and displays no correlation with the planet transit. Radiative braking is less efficient around TRAPPIST-1 than around the M2.5 dwarf GJ\,436 (\citealt{Bourrier2015_GJ436}), because its radiation pressure is about three times lower and thus more than five times lower than stellar gravity (B17). This could lead to the formation of giant hydrogen exospheres around the TRAPPIST-1 planets even larger than the one surrounding the warm Neptune GJ\,436b (\citealt{Ehrenreich2015}), and possibly extending both behind and ahead of the planets because of gravitational shear (\citealt{Bourrier2015_GJ436}). Furthermore, the XUV spectrum of TRAPPIST-1 (Sect.~\ref{sec:XUV-irrad}) yields photoionization lifetimes for neutral hydrogen atoms ranging from $\sim$20\,h at the orbit of TRAPPIST-1b to nearly 600\,h at the orbit of TRAPPIST-1h (i.e., longer than the planetary period). Because of the very low Ly-$\alpha$ and UV emission from TRAPPIST-1, neutral hydrogen exospheres could thus extend along the entire planetary orbits, and could even cross the orbit of several planets. This not only suggests that some planets could be accreting the gas escaped from their companions, but also that a large volume of the TRAPPIST-1 system could be filled with neutral hydrogen gas, providing a possible explanation for the persistent absorption signature in Visit 4. Given that the evaporating planets sustaining this system-wide hydrogen cloud would be in different relative positions at a given epoch, the structure of the cloud and its absorption signature would be highly variable over time, which could explain why it was not detected in Visit 1. We note, though, that the hydrogen cloud would still have to be accelerated to very high velocities away from the star (possibly through charge-exchange with the stellar wind, \citealt{Holmstrom2008}, \citealt{Ekenback2010}, \citealt{Bourrier2016}) to explain the velocity range of the measured absorption.

Alternatively, this variation could have a stellar origin. This is also an intriguing possibility, because the Ly-$\alpha$ line of TRAPPIST-1 was well approximated with a Gaussian profile in previous visits (B17), and the Ly-$\alpha$ line profiles of later-type M dwarfs do not show evidence for strong asymmetries (e.g., \citealt{Bourrier2015_GJ436}, \citealt{Youngblood2016}). The intrinsic Ly-$\alpha$ line of TRAPPIST-1 might have become asymmetric in Visit 4 because of variations in the up-flows and down-flows of stellar hydrogen gas, or because of absorption by colder hydrogen gas at high altitudes in the stellar atmosphere. Filaments made of partially-ionized plasma are, for the Sun, a hundred times cooler and denser than the coronal material in which they are immersed, and can thus be optically thick in the Ly-$\alpha$ line (e.g., \citealt{Parenti2014}). The large velocities of the putative absorber of TRAPPIST-1 Ly-$\alpha$ line might indicate that we witnessed the eruption of a filament that was expelled by a destabilization of the stellar magnetic field. Such eruptions can reach large distances and velocities (between 100 to 1000\,km\,s$^{-1}$  for the Sun, e.g. \citealt{Schrijver2008}). In any case, our new observations of TRAPPIST-1 raise many questions about the physical mechanisms behind the emission of the Ly-$\alpha$ line in an ultracool dwarf. \\

Despite these unknowns, our best-fit Gaussian profile can be used to estimate a conservative upper limit on the total Ly-$\alpha$ irradiation of the planets at the epoch of Visit 4 (Sect.~\ref{sec:XUV-irrad}). More observations of TRAPPIST-1 at Ly-$\alpha$ will be required to beat down the photon noise, to assess the effects of stellar variability, and to reveal absorption signatures caused by the putative planets' exospheres. The long-term variability of TRAPPIST-1 Ly-$\alpha$ line emphasizes the need for contemporaneous observations obtained outside and during the planet's transits.

\section{Evolution of the planets under high-energy irradiation}
\label{sec:evol}

Two types of spectral radiation are involved in the escape of water from an exoplanet: far-UV (FUV, 100--200~nm) to photo-dissociate water molecules and XUV (0.1--100~nm) to heat up the upper atmosphere and allow for the escape of the photo-dissociation products, hydrogen and oxygen (e.g., \citealt{VM2003}, \citealt{Lammer2003}). In this section, we study the evolution of water loss from the TRAPPIST-1 planets, in particular during their runaway greenhouse phase (see \citealt{BarnesHeller2013, LugerBarnes2015} and \citealt{Bolmont2017} for generic brown dwarfs and M-dwarfs and \citealt{Ribas2016} and \citealt{Barnes2016} for the M-dwarf planet Proxima-b). The idea is that once a planet reaches the HZ, its water can recombine and condense. Thus the amount of water reaching the upper layers of the atmosphere would be much lower than the amount available during a runaway phase. With a low mass of $0.0802 \pm 0.0073~\Msun$, TRAPPIST-1 (stellar type M8) is just above the limit between brown dwarfs and M-dwarfs and is expected to cool down for about 1\,Gyr before reaching the Main Sequence (MS). During this initial phase all TRAPPIST-1 planets, including those in the HZ today (planets e, f and g according to \citealt{Gillon2017}), were hot enough for the water potentially delivered during the formation process to be injected in gaseous form into the atmosphere, and lost more easily (\citealt{Jura2004}, \citealt{Selsis2007}).\\

\begin{figure}[htbp!]
        \centering
        \includegraphics[width=\linewidth]{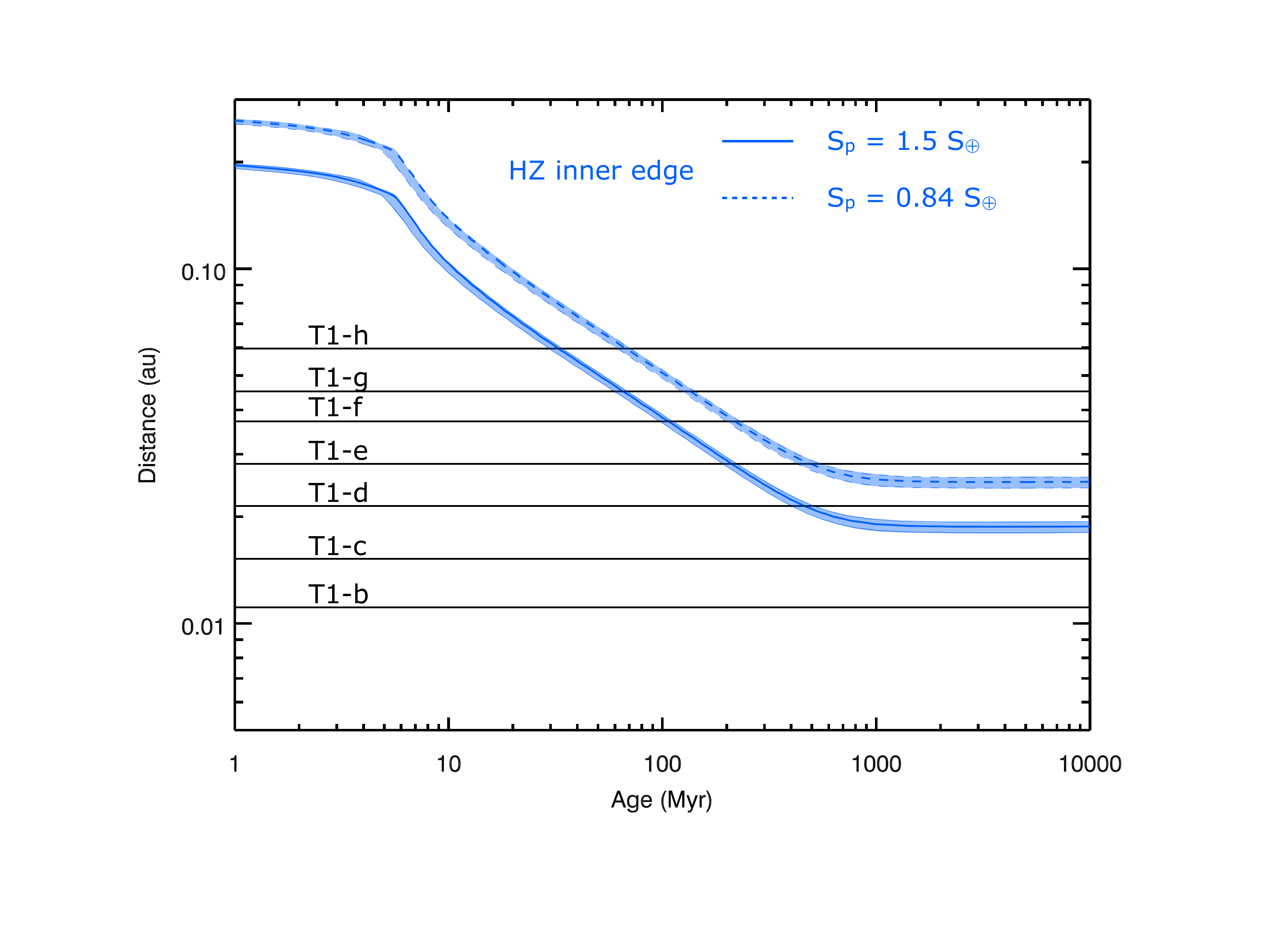}
        \caption{Architecture of the TRAPPIST-1 system and evolution of the inner edge of the HZ for two different hypotheses: a synchronized planet ($\Sp = 1.5~\Searth$, see \citealt{Yang2013}) and a non-synchronized planet ($\Sp = 0.84~\Searth$, see \citealt{Kopparapu2013}). The thick blue line corresponds to HZ inner edges which were calculated from evolutionary models for a $0.091~\Msun$ dwarf (Sect.~\ref{sec:evol} and Van Grootel et al. (submitted)). The blue areas correspond to the uncertainties on the HZ inner edge due to the uncertainty of the mass of the star.}
        \label{trappist_HZin_00875_Msun_all7planets}
        \end{figure}

In a first step we estimated the duration of the runaway greenhouse phases for the TRAPPIST-1 planets. Figure~\ref{trappist_HZin_00875_Msun_all7planets} shows the present day orbital distances of the seven planets, and the evolution of the HZ inner limits for a TRAPPIST-1 analog. Using the mass of TRAPPIST-1 ($0.0802~\Msun$) yielded an HZ inner edge much closer-in to what is shown in \citet{Gillon2017}, because low-mass star evolution models tend to underestimate the luminosity for active stars \citep{Chabrier2007} such as TRAPPIST-1 (\citealt{Luger2017}, \citealt{Vida2017}). We therefore revised the stellar mass for TRAPPIST-1 following the prescription of \citet{Chabrier2007}, by greatly reducing convection efficiency in CLES stellar evolution models (\citealt{Scuflaire2008}). We estimate a stellar mass of 0.091$\pm$0.005\,$\Msun$ using TRAPPIST-1's luminosity (\citealt{Filippazzo2015}), density and metallicity (\citealt{Gillon2017}) as inputs. The error bars include uncertainties associated with these input parameters, as well as on the initial helium abundance. This stellar mass is fully consistent with the most recent dynamical mass estimates based on ultra cool binaries for TRAPPIST-1’s spectral type (\citealt{Dupuy2017}), and with the larger radius derived by \citet{Burgasser2017}. The inferred age is greater than 2\,Gyr (the star evolves too slowly to constrain its age through stellar evolution models), consistent with the 3-8\,Gyr (\citealt{Luger2017}) and 7.6$\pm$2.2\,Gyr (\citealt{Burgasser2017}) age estimates for TRAPPIST-1. The evolution of the HZ inner edge was calculated for two different assumptions regarding the rotation of the planets. The first one ($\Sp = 1.5~\Searth$, where $\Sp$ is the insolation received by the planet and $\Searth$ the solar insolation received by the Earth) corresponds to a synchronized planet. This estimation comes from \citet{Yang2013}, which showed that a tidally locked planet could sustain surface liquid water closer to the star due to the protection of the substellar point by water clouds. The second one ($\Sp = 0.84~\Searth$) is the classical limit, computed for a non-synchronous planet \citep{Kopparapu2013}. Note that the planet does not require as much incident flux as the Earth to maintain the same surface temperature because of the redness of the star. For instance, the albedo of ice and snow is significantly lower in the infrared (\citealt{Joshi2012}), which means that the temperature of a planet is higher around TRAPPIST-1 than around a Solar-type star for a given flux and therefore, the inner edge of the HZ corresponds to a lower incoming flux. We estimated the age at which the planets entered the shrinking HZ (Table~\ref{tab_HZ_time}), considering that their migration stopped when the gas disk dissipated (\citealt{Luger2017}, \citealt{Tamayo2017}). This age corresponds to the end of the runaway greenhouse phase, and we found it lasted between a few $10$~Myr (for planet h) to a few $100$~Myr (for planet d, in the synchronized scenario). The HZ stabilized at about 1\,Gyr, earlier than the lower limit on the age of TRAPPIST-1 given by \citet{Luger2017}. This suggests that planets d to h have already been subjected to the strongest phases of their atmospheric erosion, but that planets b and c might still be in their runaway greenhouse phase if they were formed with enough water. \\
 
\begin{table*}[htbp]
\begin{center}
\caption{Age at which the TRAPPIST-1 planets enter the HZ for two different hypothesis: a synchronized planet ($\Sp = 1.5~\Searth$) and a non-synchronized planet ($\Sp = 0.84~\Searth$). }
\vspace{0.1cm}
\begin{tabular}{|c|c|c|c|c|c|c|c|}
\hline
Assumption	& \multicolumn{7}{|c|}{$T_{\rm HZ}$ (Myr)} \\
\hhline{~-------}
on HZ limit	& planet b	& planet c	& planet d	& planet e	& planet f & planet g & planet h \\
\hline
$\Sp = 1.5~\Searth$	& --	& --	& 461	& 211	& 107	& 65 		& 33 \\
$\Sp = 0.84~\Searth$	& --	& --	& -- 		& 494	& 218	& 135  	& 67 \\
\hline
\end{tabular} 
\label{tab_HZ_time} 
\end{center}
\end{table*}


\subsection{Water loss model}
\label{model}

We calculated mass-loss rates from TRAPPIST-1 planets using an improved formalism based on the energy-limited formula (e.g. \citealt{Lecav2007}, \citealt{Selsis2007}, \citealt{LugerBarnes2015}):
\begin{equation}
\label{eq:H_esc_rate}
\dot{\mathrm{M}}^{\mathrm{tot}}= \epsilon \, (\frac{R_\mathrm{XUV}}{R_\mathrm{p}})^2 \, \frac{3 \, F_\mathrm{XUV}(\mathrm{a_\mathrm{p}})}{4 \, G \, \rho_\mathrm{p} \, K_{tide}},  
\end{equation}
with $F_\mathrm{XUV}(\mathrm{a_\mathrm{p}})$ the XUV irradiation. The heating efficiency $\epsilon$ is the fraction of the incoming energy that is transferred into gravitational energy through mass loss. As in \citet{Bolmont2017} and \citet{Ribas2016}, we calculated $\epsilon$ using 1D radiation-hydrodynamic mass-loss simulations based on \citet{Owen2016}. $\epsilon$ varies with the incoming XUV radiation. For example, for today's estimated XUV flux (see next Section and Table \ref{tab_fluxes}) we obtain values of 0.064, 0.076, 0.089, 0.099, 0.107, 0.112 and 0.115, respectively for planets b to h. These efficiencies are on the same order as the 10\% assumed by \citealt{Wheatley2016} but larger than the 1\% assumed by \citet{Bourrier2017}. \\
In the following sections, we consider a constant XUV flux and an evolving XUV flux. For the latter assumption, $\epsilon$ is computed accordingly. In the energy-limited formula the parameter $(\frac{R_\mathrm{XUV}}{R_\mathrm{p}})^2$ accounts for the increased cross-sectional area of planets to XUV radiation, while $K_\mathrm{tide}$ accounts for the contribution of tidal forces to the potential energy (\citealt{Erkaev2007}). Both are set to unity for these cool and small planets (\citealt{Bolmont2017}). The mean density of the planets was calculated using masses derived from TTV in \citealt{Gillon2017} (See Table~\ref{tab_trapp}). We consider here that the atmospheres of the planets are mainly composed of hydrogen and oxygen and compute their joint escape using the formalism of \citet{Hunten1987}. The escape rates of both elements depend on the temperature of the thermosphere, the gravity of the planet and a collision parameter between oxygen and hydrogen. As in \citet{Bolmont2017} and \citet{Ribas2016}, we adopt a thermosphere temperature of 3000~K obtained through our hydrodynamic simulations. We caution that more detailed models, including FUV radiative transfer, photochemical schemes, and non-LTE kinetics in the rarefied gas regions of the upper atmosphere, will be required to determine accurately the outflow properties. For example the hydrodynamic outflow of hydrogen could drag water molecules, which are only slightly heavier than oxygen atoms, upward and they would be photodissociated at high altitudes into more escaping oxygen and hydrogen atoms. Nonetheless, our assumptions likely maximize the XUV-driven escape (see \citealt{Bolmont2017}), and our estimates of the water loss should be considered as upper limits. 


\subsection{Estimation of the planets' XUV irradiation}
\label{sec:XUV-irrad}

To calculate the planetary mass losses, we needed estimations of the XUV irradiation from TRAPPIST-1 over the whole history of the system. In a first step, we calculated the present day stellar irradiation. We used the same value as in \citet{Bourrier2017} for the X-ray emission (5--100\,\AA\,), studied by \citet{Wheatley2016}. The stellar EUV emission between 100--912\,\AA\, is mostly absorbed by the ISM but can be approximated from semi-empirical relations based on the Ly-$\alpha$ emission. The theoretical Ly-$\alpha$ line profile derived for Visit 4 (Sect.~\ref{sec:cor_struc}) yields an upper limit on the total Ly-$\alpha$ emission of 7.5$\pm$0.9$\times$10$^{-2}$\,erg\,s$^{-1}$\,cm$^{-2}$ at 1\,au. This is larger than the emission derived for previous visits (5.1$\stackrel{+1.9}{_{-1.3}}\times$10$^{-2}$\,erg\,s$^{-1}$\,cm$^{-2}$, B17), in agreement with the increase in flux suggested by observations (Sect.~\ref{sec:long_term}). We chose to consider those two estimates as lower and upper limits on the present Ly-$\alpha$ emission of TRAPPIST-1, and used the \citet{Linsky2014} relation for M dwarfs to derive corresponding limits on the EUV flux. Table~\ref{tab_fluxes} gives our best estimate for today's range of fluxes emitted by TRAPPIST-1 at Ly-$\alpha$ and between 5 - 912\,\AA. We computed the ratio of these two fluxes to the bolometric luminosity and obtained a value of log$_{10}(\Lxuv/L_{\rm bol})$ between -3.39 and -3.73, which is about a factor 2.5 lower than estimated from the X-ray flux by \citet{Wheatley2016}. \\

In a second step, we estimated the past stellar irradiation. We consider that when the planets were embedded in the protoplanetary disk, they were protected from irradiation and did not experience mass loss. In that frame, water loss began at the time when the disk dissipated, which we assume to be 10~Myr \citep{Pascucci2009, Pfalzner2014, PecautMamajek2016}. We investigated two different scenarios, depending on our assumptions for the temporal evolution of the XUV emission after the disk dissipation:
\begin{enumerate}
\item[-] A constant $\Lxuv$ equal to today's range of emission. This assumption might be supported by the X-ray flux of TRAPPIST-1, which is consistent with a saturated emission typical of earlier type M-dwarfs, according to \citealt{Wheatley2016}).
\item[-] An evolving $\Lxuv$, considering the ratio $\Lxuv/L_{\rm bol}$ to be constant throughout the history of the star. The ratio was set to the present day estimate of the star luminosities (see Table \ref{tab_fluxes}). We used the evolutionary models of \citet{Baraffe2015} to compute the evolution of the bolometric luminosity.
\end{enumerate}
Here we set the stellar mass to its nominal value of 0.091~$\Msun$. Considering the range allowed by the uncertainties would slightly change our water loss estimates, at the time the planet reached the HZ for the constant XUV flux prescription, and at all ages for the evolving XUV flux prescription. We estimate that the difference would be less than 4\% at an age of 8~Gyr.

\begin{table*}[htbp]
\begin{center}
\caption{High-energy emission from TRAPPIST-1.}
\vspace{0.1cm}
\begin{tabular}{|c|c|c|c|c|}
\hline
Wavelength domain 	& \multicolumn{2}{|c|}{XUV ($0.5 - 100$ nm)} 				& \multicolumn{2}{|c|}{$L_\alpha$}  \\
\hhline{~----}

				& $L_{\rm XUV}$ (erg.s$^{-1}$)	& log$_{10}(L_{\rm XUV}/L_{\rm bol})$ 	& $L_{\alpha}$ (erg.s$^{-1}$)	& log$_{10}(L_\alpha/L_{\rm bol})$ \\
\hline
Lower estimate 	& $5.26\times10^{26}$ 	& -3.58		& $1.44\times10^{26}$	& -4.15 \\
Mean estimate		& $6.28\times10^{26}$	& -3.51		& $1.62\times10^{26}$	& -4.09 \\
Upper estimate 	& $7.30\times10^{26}$ 	& -3.44		& $1.81\times10^{26}$	& -4.05 \\
\hline
\end{tabular} 
\label{tab_fluxes} 
\end{center}
\end{table*}


\subsection{Water loss evolution}
\label{sec:wat_loss_ev}

\begin{figure*}[ht!]
     	\begin{center}
        	\subfigure[Constant XUV luminosity]{%
            \label{Lxcst_lum}
            \includegraphics[width=0.35\textwidth]{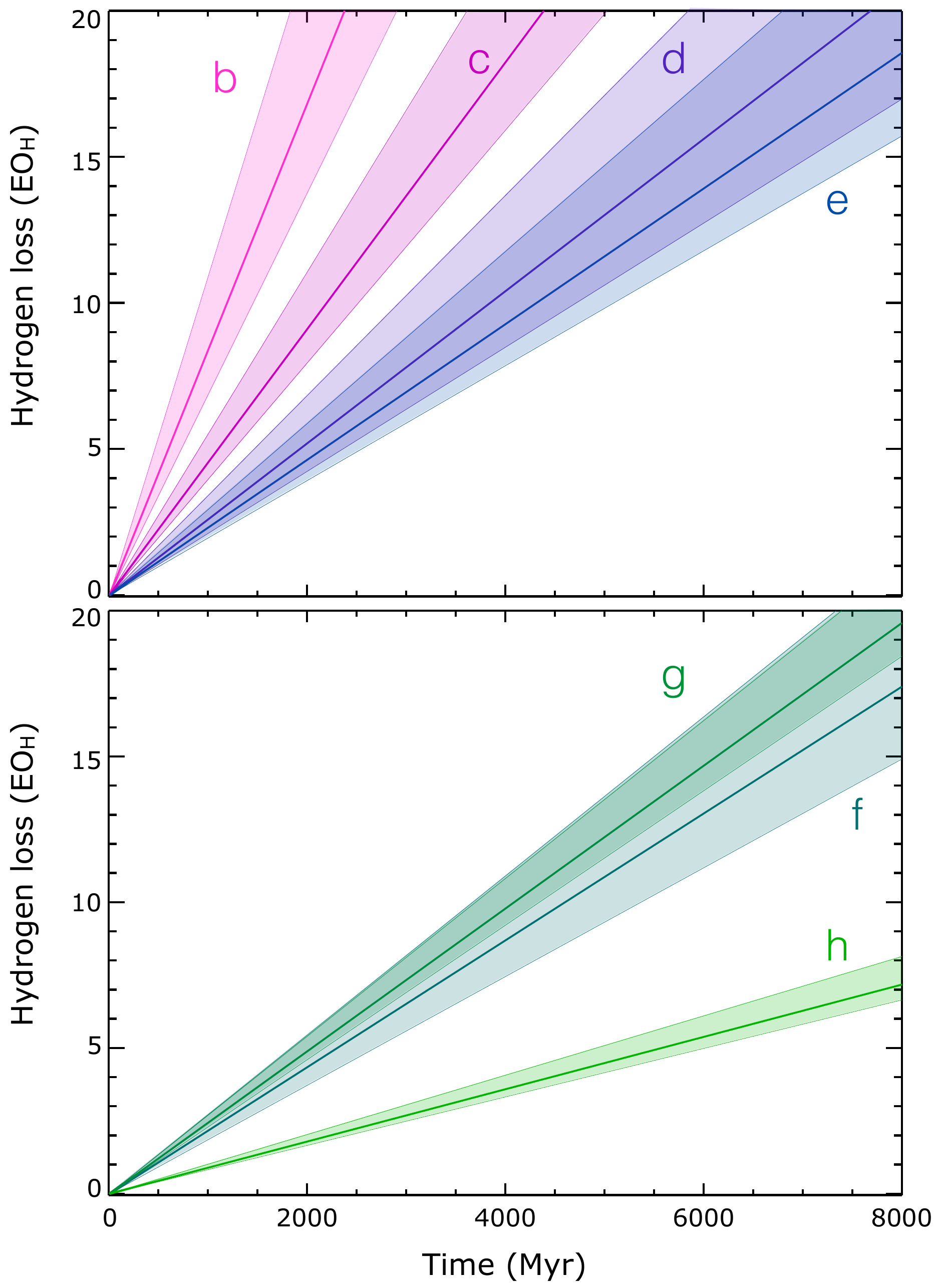}
        } \hspace{0cm}
        \subfigure[Evolving XUV luminosity]{%
           \label{Levol_lum}
           \includegraphics[width=0.35\textwidth]{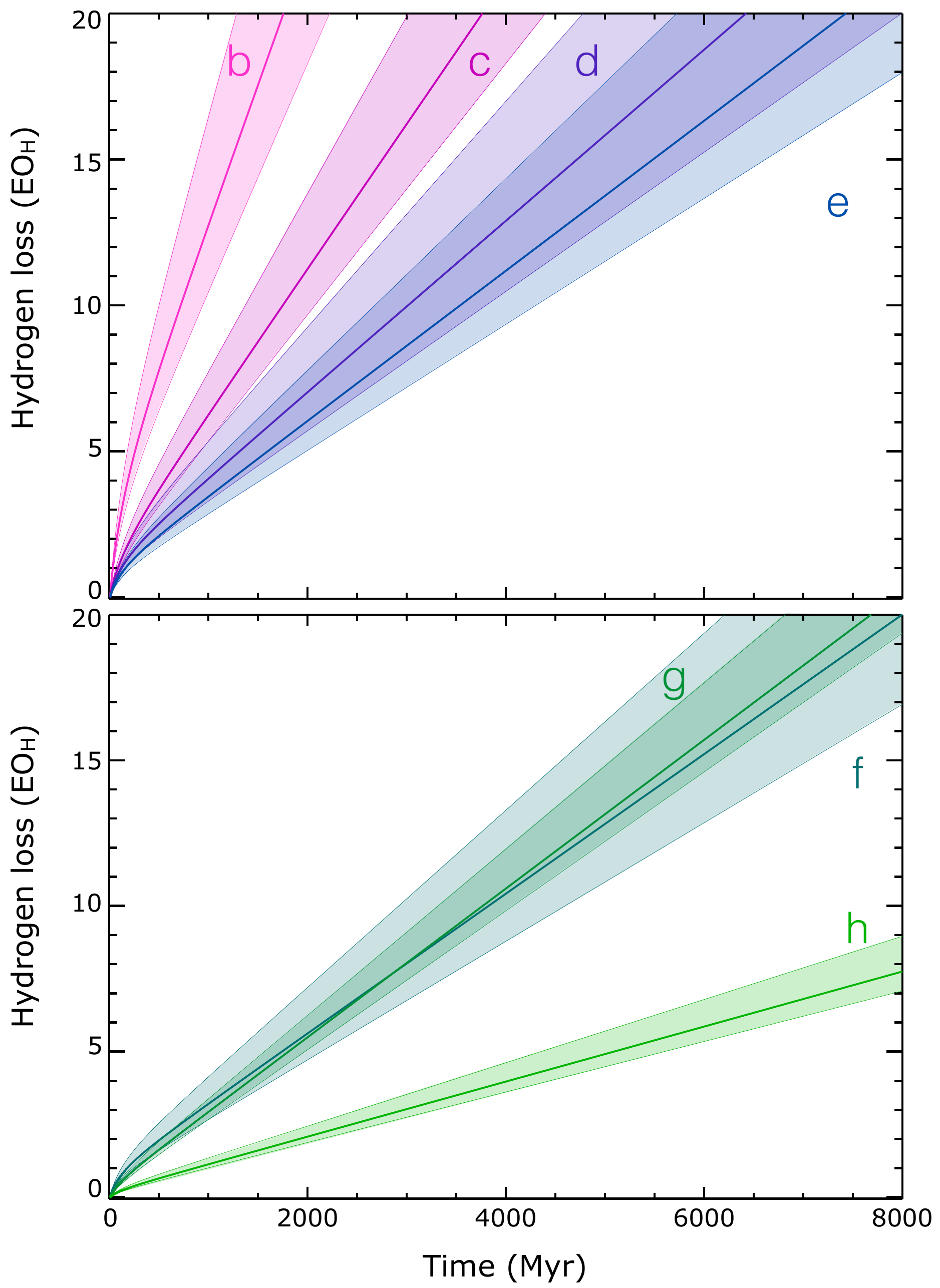} 
        }
    	\end{center}
    	\caption{Cumulative hydrogen loss for the TRAPPIST-1 planets for the two different assumptions for the evolution and values of the XUV flux (see Table~\ref{tab_fluxes}). The masses of the planets are the masses from Table~\ref{tab_trapp}. Because of our assumptions, these estimates likely represent upper limits on the actual loss.}
   	\label{waterloss_diff_lum}
	\end{figure*}

Using the energy-limited model in Sect.~\ref{model} and our estimates for the XUV irradiation in Sect.~\ref{sec:XUV-irrad}, we calculated the mass loss from TRAPPIST-1 planets over time. In order to calculate the hydrogen loss, we used the method (2) of \citet{Ribas2016}, which consists in calculating the ratio between the oxygen and hydrogen fluxes as a function of the XUV luminosity. We consider here an infinite water reservoir. This allows us to consider that the ratio of hydrogen and oxygen remains stoichiometric even though the loss is not stoichiometric. This provides us with an upper limit on the mass loss (see \citealt{Ribas2016} for a discussion on the effect of a finite initial water reservoir). The loss is given in units of Earth Ocean equivalent content of hydrogen (referred as $1~EO_H$). In other words, the mass loss is expressed in unit of the mass of hydrogen contained in one Earth ocean ($1.455\times 10^{20}$~kg, with an Earth ocean mass corresponding to 1.4$\times$10$^{21}$\,kg). For example we estimate a current mass loss from planet b of $0.008~M_{\rm ocean}$/Myr, for the nominal XUV flux (Table~\ref{tab_fluxes}). This corresponds to escape rates of oxygen and hydrogen from planet b of $2.9\times10^8$~g/s and $4.3\times10^7$~g/s, respectively. The values for the other planets can be found in Table~\ref{tab_rates}. Figure~\ref{waterloss_diff_lum} shows the evolution of the hydrogen loss from the TRAPPIST-1 planets as a function of the age of the system in the two scenarios assumed for the evolution of the XUV flux. Table~\ref{waterlost1} gives the corresponding mass loss at different times of interest. \\

\begin{table*}[htbp]
\begin{center}
\caption{Current mass loss rate, hydrogen loss rate and oxygen loss rate for the TRAPPIST-1 planets.}
\vspace{0.1cm}
\begin{tabular}{|c|c|c|c|c|c|c|c|}
\hline
Planet						& b				& c 		 		& d				& e				& f				& g				& h		\\
\hline
Mass loss ($M_{\rm ocean}/$Myr)	&$8.2\times10^{-3}$ &$2.9\times10^{-3}$ 	&$2.3\times10^{-3}$ 	&$1.7\times10^{-3}$ 	&$1.4\times10^{-3}$ 	&$6.4\times10^{-4}$ 	&$2.9\times10^{-4}$ \\ 
Hydrogen loss (g/s)				&$4.3\times10^{7}$  &$2.3\times10^{7}$  	&$1.3\times10^{7}$  	&$1.2\times10^{7}$  	&$1.1\times10^{7}$  	&$1.2\times10^{7}$  	&$4.3\times10^{6}$  \\  
Oxygen loss (g/s) 				&$2.9\times10^{8}$  &$1.0\times10^{8}$	&$8.2\times10^{7}$	&$5.7\times10^{7}$	&$4.7\times10^{7}$	&$1.5\times10^{7}$	&$7.5\times10^{6}$  \\	
\hline
\end{tabular} 
\label{tab_rates} 
\end{center}
\end{table*}

Planet b and c never reach the HZ, so they are expected to have lost water via the runaway greenhouse mechanism (Sect.~\ref{sec:evol}) throughout the full lifetime of the star. If TRAPPIST-1 is 3~Gyr old (the lower estimate of the age given in \citealt{Luger2017}), planet b potentially lost more than 20~$EO_H$ and planet c more than $10~EO_H$. If planets b and c formed with an Earth-like water content, they are likely dry today, whatever the assumptions on the XUV flux and the age of the star. Alternatively they might have formed as ocean planets (\citealt{leger2004}), in which case a loss of 20~Earth oceans for planet b would represent only 0.5\% of its mass. Currently this scenario is not favoured by the planet formation model proposed by \citet{Ormel2017}, which excludes water fractions larger than about 50\%, and by the densities of planets b and c derived observationally by \citet{Gillon2017} and \citet{Wang2017}, although we note that they still allow for a significant water content. \\

If we consider that the water loss only occurs during the runaway phase, planets d to f lost less than $4~EO_H$ before reaching the HZ, and planets g and h lost less than $1~EO_H$. In that scenario the outer planets of the TRAPPIST-1 system might thus still harbour substantial amounts of water, especially planets e to h if the low densities derived by \citet{Wang2017} are confirmed. What if hydrodynamic water loss continued once the planets reached the HZ? After 3\,Gyr, we estimate that planets g and those closer-in would have lost more than $7~EO_H$. After 8\,Gyr, they would have lost more than $20~EO_H$ (Table~\ref{waterlost1}). Interestingly, the relation obtained by \citealt{Guinan2016} for M0-5 V dwarf stars yields an age of about 7.6\,Gyr for TRAPPIST-1, using the X-ray flux obtained by \citealt{Wheatley2016} in the ROSAT band (0.14\,erg\,s$^{-1}$\,cm$^{-2}$ at 1\,au). This age is at the upper limit of the range derived by \citet{Luger2017}, and if confirmed suggests that all TRAPPIST-1 planets have lost substantial amounts of water over the long history of the system. Refined estimates of the planet densities will however be necessary to determine whether they still harbour a significant water content. We also note that our estimates for the water loss once planets are in the HZ (Table~\ref{waterlost1}) are probably upper limits. If the planet is able to retain its background atmosphere, the tropopause is expected to act as an efficient cold trap, preventing water from reaching the higher parts of the atmosphere \citep[e.g.,][]{WordsworthPierrehumbert2014, Turbet2016}. In that case, water escape would be limited by the diffusion of water through the cold trap. However, if the background pressure is low, the water vapor mixing ratio increases globally in the atmosphere \citep{Turbet2016} and hydrogen escape is no longer limited by the diffusion of water. The farther out the TRAPPIST-1 planets, the more likely there would have been able to sustain an important background atmosphere, thus protecting the water reservoir once in the HZ.

\subsection{Consequences of the uncertainty on the planetary masses}

\begin{figure*}[ht!]
     	\begin{center}
        	\subfigure[Effect of mass on the hydrogen flux]{%
            \label{effect_mass}
            \includegraphics[width=0.35\textwidth]{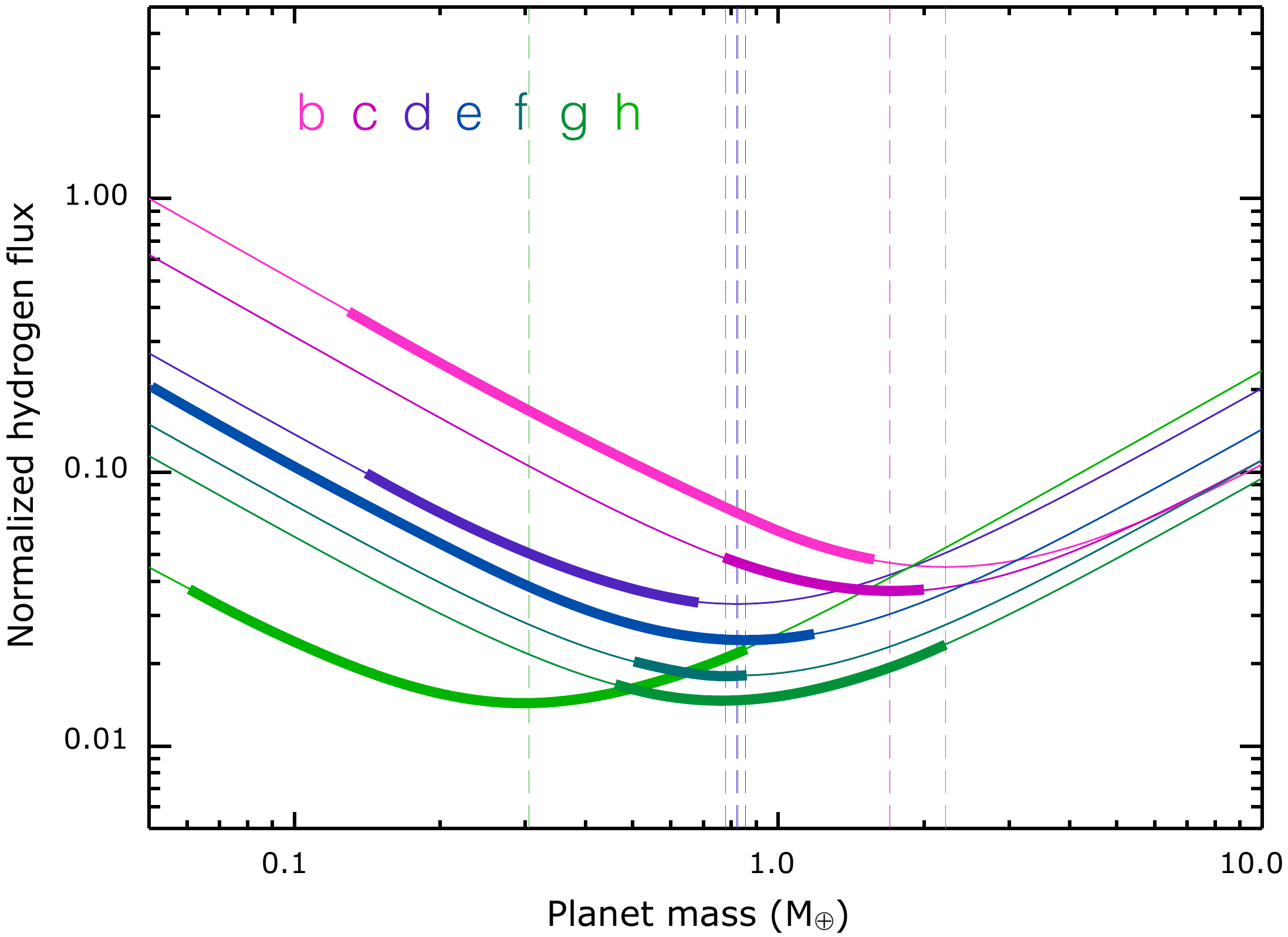}
        } \hspace{0cm}
        \subfigure[Hydrogen loss for the observational mass range]{%
           \label{Levol_mass}
           \includegraphics[width=0.35\textwidth]{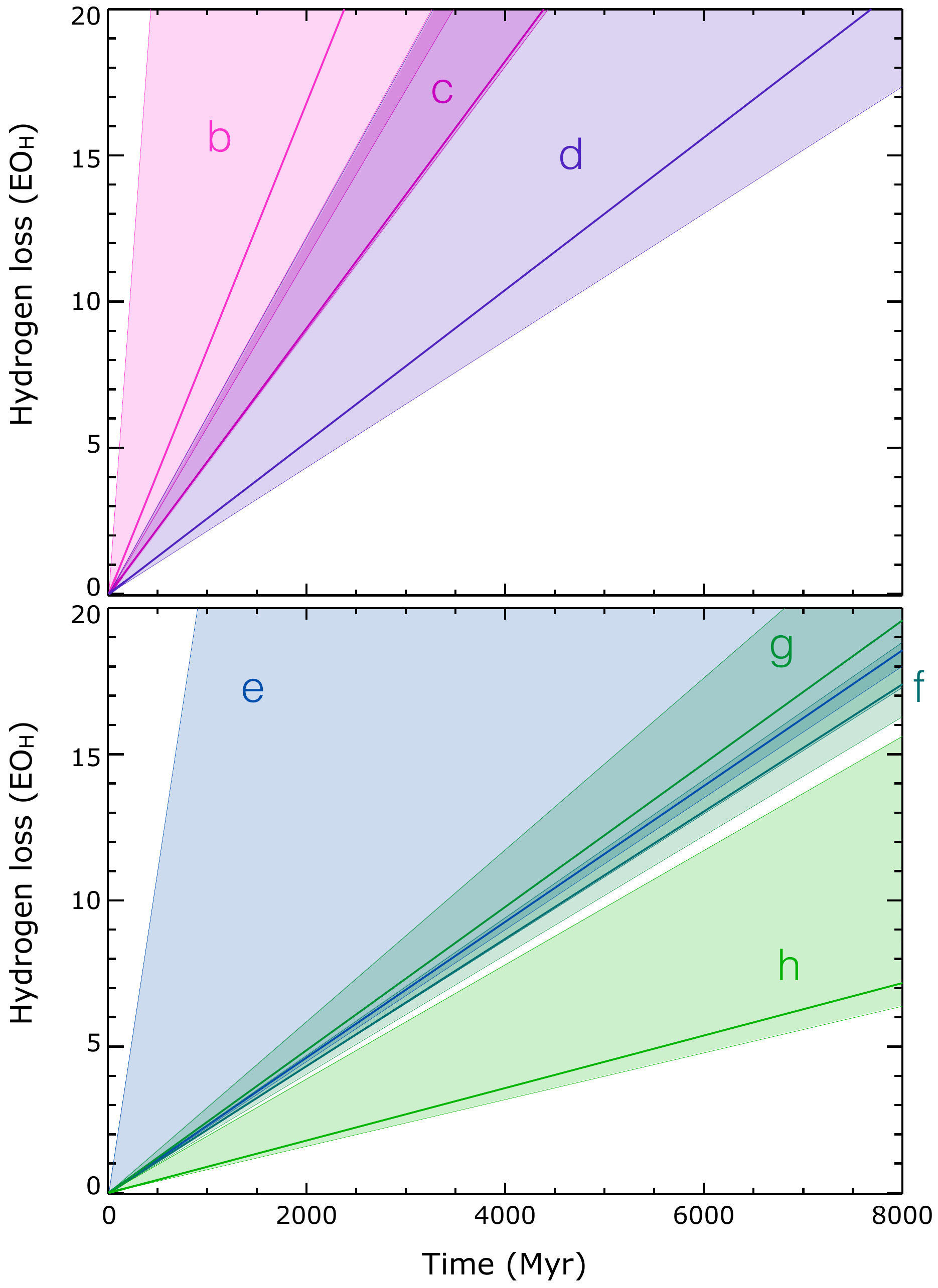} 
         }
    	\end{center}
    	\caption{Effect of planetary mass on the hydrogen loss for the TRAPPIST-1 planets, for the range of masses in \citet{Gillon2017}. The XUV flux is set to its nominal estimate ($L_{\rm XUV}$ = cst) (a) Normalized hydrogen flux as a function of planetary mass for the different planets. The thick part of each curve corresponds to the allowed range determined by \citet{Gillon2017} and the vertical line corresponds to the mass which corresponds to the lowest hydrogen flux. (b) Hydrogen loss from the planets. The thicker line corresponds to the loss calculated for the mass given in \citet{Gillon2017} and the thin lines correspond to the lower and upper estimation within the mass range given in the same article.}
   	\label{waterloss_diff_mass}
	\end{figure*}

While the radii of the TRAPPIST-1 planets are known to a good precision, their masses remain very uncertain. We set today's XUV luminosity to the nominal estimate in Table~\ref{tab_fluxes}, and investigated the effect of changing the value of the planetary mass within the range of uncertainty estimated by \citet{Gillon2017} (see Table~\ref{tab_trapp}). For planet h, we investigated a range of compositions from $100$\% ice to 100\% iron, which corresponds to masses between $0.06~\Mearth$ and $0.86~\Mearth$. Figure~\ref{Levol_mass} shows that for most planets the uncertainty on their mass dominates the final uncertainty on the mass loss, compared to the effect of varying the XUV irradiation with its uncertainty (see also \citealt{LugerBarnes2015}). This is because hydrogen loss is not only inversely proportional to the planetary mass (see Eq. 1), but is also linked to the ratio of the escape fluxes of hydrogen and oxygen ($r_{\rm F} = F_{\rm O}/F_{\rm H}$), which is a function of the cross-over mass \citep[see][]{Hunten1987, Bolmont2017}. For a given value of the XUV luminosity and an infinite initial water reservoir, the hydrogen mass loss is described by a polynomial in $\Mp$ of the form $\alpha/\Mp+\beta\Mp$, where $\alpha$ and $\beta$ are constants depending, for example, on the radius of the planet (we refer to Eqs. 8 and 9 of \citealt{Bolmont2017} to derive these values). The mass flux decreases with increasing planetary mass and the ratio of the flux of oxygen over the flux of hydrogen decreases with increasing planetary mass. For high masses, the mass flux is lower (gravity wins over cross section) but the mass loss mainly occurs via hydrogen escape. This results in a high hydrogen escape rate. For low masses, the outflow is a mixture of hydrogen and oxygen, meaning less hydrogen escape for an equal energy input, but the overall mass flux is higher. For low enough mass, the increased overall mass flux does more than compensate and this also results in a high hydrogen loss. The hydrogen loss is therefore high for very low mass and very high mass planets with a minimum for intermediate mass. Figure~\ref{effect_mass} shows this dependance of the hydrogen flux with mass for each planet of the system. For all planets but planet b, the minimum loss is achieved for an intermediate mass in the range allowed by the observations (this range is displayed as a thicker line on the graph). The lower curves delimiting the colored areas in Figure~\ref{Levol_mass} correspond to the hydrogen loss calculated for this intermediate mass for all planets but b. For planet b, the minimum hydrogen loss is obtained for the highest mass allowed by the observations. \\

In the end the uncertainty on the hydrogen loss comes mainly from the uncertainty on mass rather than the uncertainty on the prescribed luminosity (although we note that our calculations assume an unlimited water supply, yielding upper limits on the water losses; see Sect.~\ref{sec:wat_loss_ev}). The uncertainty on the hydrogen loss thus ranges from about 80\% for planets b and e, $\sim50$\% for planets d and h, $\sim$20\% for planet c and g and only 4\% for planet f. Table~\ref{waterloss_massp} gives the upper and lower estimates of hydrogen loss for the mass range for each planet and for the different times considered in this study. Due to the relative precision of the mass of planet f, the mass loss is relatively well constrained with our model, with a loss of less than 0.5~$EO_H$ before reaching the HZ and less than $20~EO_H$ at an age of 8~Gyr. Given that the low density of planet f is compatible with a non negligible water content \citep{Gillon2017}, this could indicate that water loss may not have been a very efficient process and/or that the planet formed with a large fraction of its mass in water. \\

We note that for most planets, the estimates of the masses from \citet{Wang2017} are more precise but consistent compared to \citet{Gillon2017}. However, for planet f, the mass ranges obtained by these different studies are incompatible, and using the mass from \citet{Wang2017} would lead to a higher mass loss than what is shown on Figure~\ref{Levol_mass}. More data is needed to refine our measurements of the masses of the planets of the system. 

\begin{table*}
\centering
\caption{Cumulative hydrogen loss (in EO$_H$) for different times (table corresponding to Fig.~\ref{waterloss_diff_lum}). T$_\mathrm{HZ}$ is the age at which a planet enters in the HZ (see Table~\ref{tab_HZ_time}). The two values given 
for each column correspond to the uncertainty coming from the different luminosity prescriptions (between low and high, see Table~\ref{tab_fluxes}).}
\vspace{0.1cm}
\begin{tabular}{|c||c|c|c|c|c|c||}
\hline
\multicolumn{1}{|c||}{}  & \multicolumn{1}{|c|}{}  		& \multicolumn{1}{|c|}{}  			& \multicolumn{4}{c||}{H loss (EO$_H$)}  \\
\multicolumn{1}{|c||}{}  & \multicolumn{1}{|c|}{Planet}	& \multicolumn{1}{|c|}{Mass}  		& $T_{\rm HZ}$      	& $T_{\rm HZ}$		& 3~Gyr	& 8~Gyr   \\
\multicolumn{1}{|c||}{}  & \multicolumn{1}{|c|}{}		& \multicolumn{1}{|c|}{($\Mearth$)}  	& (1.5~S$_{\oplus}$)	& (0.84~S$_{\oplus}$)	&      		&  		\\
\hline
$L_{\rm XUV}$ evol 	& b	& 0.85 	& --				& --				& 29.2--35.4	& 71.5--86.9	\\
			 	& c	& 1.38	& --				& --				& 15.2--17.6	& 38.1--44.1	\\
			 	& d	& 0.41	& 2.35--2.87	& --				& 9.12--11.2  & 22.2--27.2 	\\
			 	& e	& 0.62	& 1.20--1.46	& 2.05--2.49	& 7.98--9.57	& 19.6--23.4 	\\
			 	& f	& 0.68	& 0.70--0.86	& 1.14--1.39	& 7.46--8.89	& 18.4--21.8 	\\
			 	& g	& 1.34	& 0.31--0.37 	& 0.58--0.67	& 7.79--8.46	& 20.1--21.7 	\\
				& h	& 0.46	& 0.07--0.09	& 0.14--0.17	& 2.91--3.23	& 7.40--8.15	\\
\hline
\hline
$L_{\rm XUV}$ cst 	& b	& 0.85 	& --				& --			& 25.3--30.7	& 67.5--82.1	\\
			 	& c	& 1.38	& --				& --			& 13.7--15.8	& 36.5--42.1	\\
			 	& d	& 0.41	& 1.17--1.44	& --			& 7.80--9.57  	& 20.8--25.5 	\\
			 	& e	& 0.62	& 0.47--0.56	& 1.12--1.34	& 6.95--8.28	& 18.6--22.1 	\\
			 	& f	& 0.68	& 0.21--0.25	& 0.45--0.54	& 6.52--7.70	& 17.4--20.6 	\\
			 	& g	& 1.34	& 0.14--0.15 	& 0.31--0.33	& 7.34--7.89	& 19.6--21.1 	\\
				& h	& 0.46	& 0.02--0.02	& 0.05--0.06	& 2.69--2.94	& 7.18--7.86	\\
\hline
\end{tabular} 
\label{waterlost1}
\end{table*}

\begin{table*}
\centering
\caption{Cumulative hydrogen loss (in EO$_H$) for different times. T$_\mathrm{HZ}$ is the age at which a planet enters in the HZ (see Table~\ref{tab_HZ_time}). The two values given for each column correspond to the uncertainty coming from the masses (for the mean estimation of the XUV luminosity, see Table~\ref{tab_fluxes}).}
\vspace{0.1cm}
\begin{tabular}{|c||c|c|c|c|c|c|c||}
\hline
\multicolumn{1}{|c||}{}  & \multicolumn{1}{|c|}{}  		& \multicolumn{1}{|c|}{Mass}  			& \multicolumn{4}{c|}{H loss (EO$_H$)}  & Uncertainty\\
\multicolumn{1}{|c||}{}  & \multicolumn{1}{|c|}{Planet}	& \multicolumn{1}{|c|}{range}  		& $T_{\rm HZ}$      	& $T_{\rm HZ}$		& 3~Gyr	& 8~Gyr & range   \\
\multicolumn{1}{|c||}{}  & \multicolumn{1}{|c|}{}		& \multicolumn{1}{|c|}{($\Mearth$)}  	& (1.5~S$_{\oplus}$)	& (0.84~S$_{\oplus}$)	&      		&  	& \%	\\
\hline
$L_{\rm XUV}$ cst 	& b	& 0.13--1.57 	& --			& --			& 18.8--160	& 50.3--429	& 79	\\
			 	& c	& 0.77--1.99	& --			& --			& 14.5--19.2	& 38.6--51.3	& 14	\\
			 	& d	& 0.14--0.68	& 1.04--3.16	& --			& 6.97--21.0  	& 18.6--56.0 	& 50	\\
			 	& e	& 0.04--1.20	& 0.49--5.21	& 1.17--12.5	& 7.26--77.6	& 19.4--207 	& 83	\\
			 	& f	& 0.50--0.86	& 0.23--0.25	& 0.48--0.55	& 6.99--7.88	& 18.7--21.0 	& 6	\\
			 	& g	& 0.46--2.22	& 0.12--0.19	& 0.28--0.43	& 6.60--10.3	& 17.6--27.5 	& 22	\\
				& h	& 0.06--0.86	& 0.02--0.05	& 0.05--0.13	& 2.59--6.84	& 6.92--18.3	& 45	\\
\hline
\end{tabular} 
\label{waterloss_massp}
\end{table*}


\section{Hydrogen loss vs hydrogen production}
\label{sec:h_loss_crea}
       
\subsection{Photolysis}       
       
\begin{figure}[htbp!]
        \centering
        \includegraphics[width=\linewidth]{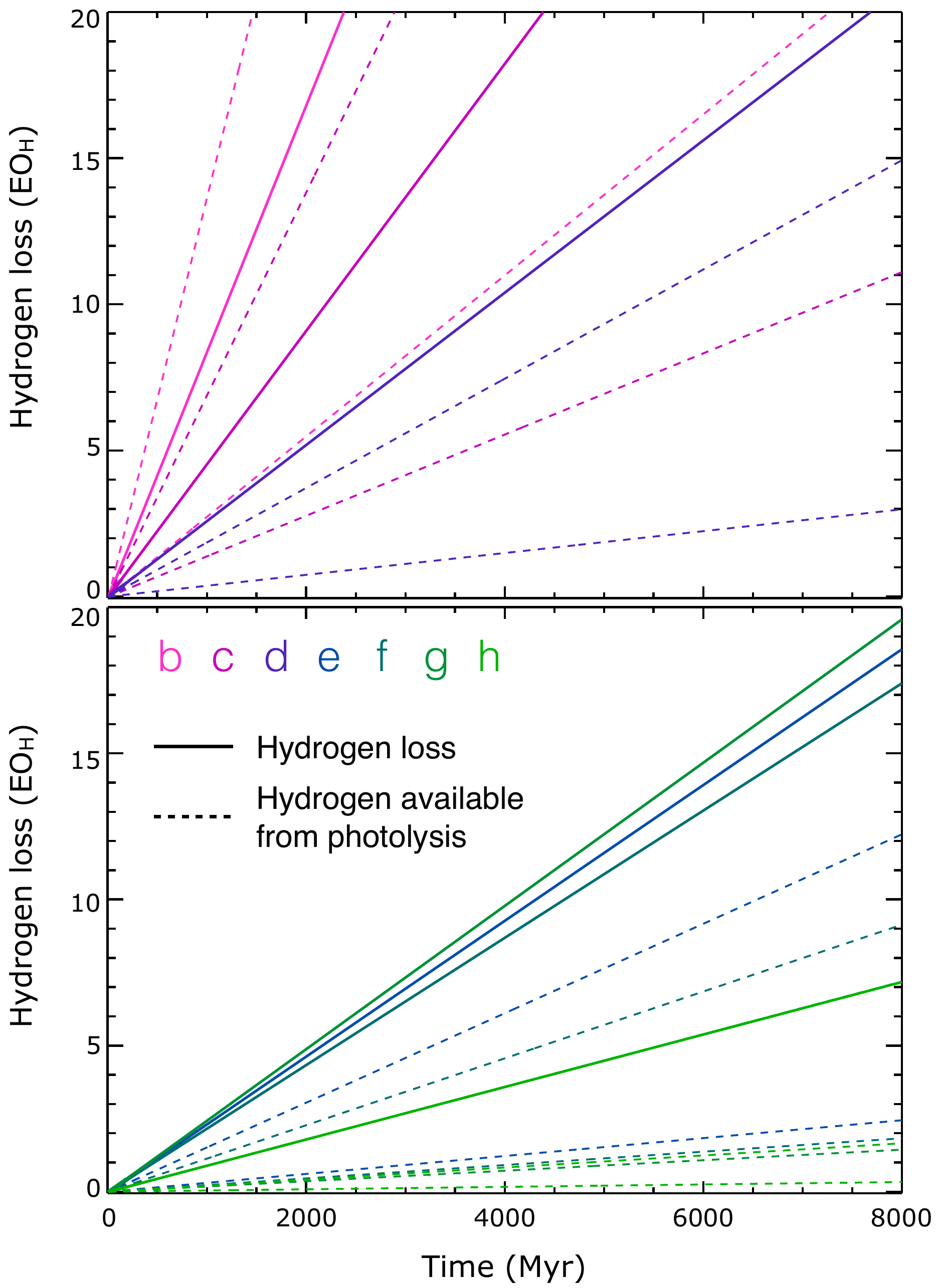}
        \caption{Hydrogen loss (full lines) and hydrogen production (dashed lines) by photolysis for the planets of TRAPPIST-1. The amount of hydrogen formed by photolysis was calculated for two efficiencies of the process: 1 (upper dashed lines) and 0.2 (lower dashed lines). With a realistic value of the photolysis efficiency, we find that photolysis is the limiting process as the hydrogen loss cannot occur faster than the hydrogen production.}
        \label{massloss_TRAPPIST_Ms_00802_tinit_10Myr_EOH_photodiss_nom123_newrf_vincent_Lx_evol}
        \end{figure}

The atmospheric mass loss can be limited by the amount of hydrogen formed by photo-dissociation of water molecules. We computed the rate of hydrogen production driven by the FUV part of the spectrum, which is taken to be restricted to the Ly $\alpha$ emission \citep[as in][]{Bolmont2017}. We note that water molecules could further be dissociated through impact with high-energy electrons, in particular those produced by the ionization of water (considering the high X-ray emission of TRAPPIST-1 and the large ionization cross-section of water in the XUV; \citealt{Heays2017}). Figure~\ref{massloss_TRAPPIST_Ms_00802_tinit_10Myr_EOH_photodiss_nom123_newrf_vincent_Lx_evol} shows the hydrogen loss for the nominal irradiation (Table~\ref{tab_fluxes}) and planet masses, and the hydrogen quantity available due to photo-dissociation of water for two different efficiencies:  $\epsilon_\alpha =1$ (each photon leads to a dissociation, see \citealt{Bolmont2017}) and 0.2. We obtained the following results for a constant XUV luminosity:
\begin{enumerate}
\item For planets b and c, photo-dissociation is not the limiting process for high efficiencies. The hydrogen loss is limited by the hydrodynamic escape as computed in the previous section;
\item For planets d to g, photo-dissociation is the limiting process whatever the efficiency: the rate of hydrogen formation by photo-dissociation is below the escape rate of hydrogen.
\end{enumerate}
Photo-dissociation of water also becomes the limiting process for planets b and c if $\epsilon_\alpha \approxinf 0.60$. Because of various processes, such as photon backscattering or recombination of hydrogen atoms, only a fraction of the incoming FUV photons actually results in the loss of a hydrogen atom. As a result we do not expect photo-dissociation to be more efficient than 20\% (\citealt{Bolmont2017}). It should be the limiting process for all planets, and the mass losses estimated in Sect.~\ref{sec:evol} can be considered as upper limits. The quantity of hydrogen avaible from photolysis assuming a 20\% efficiency is given in Table~\ref{waterloss_photodisss} for each planet, to be compared with the hydrodynamic hydrogen loss. We note that the photolysis process does not depend on the mass of the planets (only on their radii, known to a high precision for the TRAPPIST-1 planets), so these results are more robust than the hydrogen loss estimates, which highly depend on the mass of the planets. 

\begin{table*}
\centering
\caption{Hydrogen loss and hydrogen production (in EO$_H$) for different times. The two values given for each column correspond to the quantity of hydrogen lost (calculated for the nominal estimate of the XUV luminosityin Table~\ref{tab_fluxes}, and the nominal masses) and the quantity of hydrogen available from photolysis (in bold, assuming an efficiency of 20\%).}
\vspace{0.1cm}
\begin{tabular}{|c||c|c|c|c|c|c||}
\hline
\multicolumn{1}{|c||}{}  & \multicolumn{1}{|c|}{}  		& \multicolumn{1}{|c|}{}  			& \multicolumn{4}{c||}{H loss (EO$_H$)}  \\
\multicolumn{1}{|c||}{}  & \multicolumn{1}{|c|}{Planet}	& \multicolumn{1}{|c|}{Mass}  		& $T_{\rm HZ}$      	& $T_{\rm HZ}$		& 3~Gyr	& 8~Gyr   \\
\multicolumn{1}{|c||}{}  & \multicolumn{1}{|c|}{}		& \multicolumn{1}{|c|}{($\Mearth$)}  	& (1.5~S$_{\oplus}$)	& (0.84~S$_{\oplus}$)	&      		&  		\\
\hline
$L_{\rm XUV}$ evol 	& b	& 0.85 	& --				& --								& 32.4--\textbf{13.0}	& 79.5--\textbf{28.6}	\\
			 	& c	& 1.38	& --				& --								& 16.5--\textbf{6.54}	& 41.2--\textbf{14.4}	\\
			 	& d	& 0.41	& 2.62--\textbf{0.65}	& --						& 10.2--\textbf{1.76}  	& 24.8--\textbf{3.88} 	\\
			 	& e	& 0.62	& 1.33--\textbf{0.39}	& 2.28--\textbf{0.55}	& 8.79--\textbf{1.44}	& 21.6--\textbf{3.18} 	\\
			 	& f	& 0.68	& 0.79--\textbf{0.21}	& 1.27--\textbf{0.29}	& 8.19--\textbf{1.08}	& 20.1--\textbf{2.38}	\\
			 	& g	& 1.34	& 0.34--\textbf{0.12}	& 0.63--\textbf{0.19}	& 8.13--\textbf{0.85}	& 20.9--\textbf{1.87} 	\\
				& h	& 0.41	& 0.08--\textbf{0.02}	& 0.16--\textbf{0.03}	& 3.07--\textbf{0.20}	& 7.78--\textbf{0.43}	\\
\hline
\hline
$L_{\rm XUV}$ cst 	& b	& 0.85 	& --					& --					& 28.1--\textbf{9.30}	& 75.0--\textbf{24.8}	\\
			 	& c	& 1.38	& --					& --					& 14.8--\textbf{4.69}	& 39.4--\textbf{12.5}	\\
			 	& d	& 0.41	& 1.31--\textbf{0.19}		& --					& 8.70--\textbf{1.26}  & 23.2--\textbf{3.37} 	\\
			 	& e	& 0.62	& 0.51--\textbf{0.07}		& 1.23--\textbf{0.17}		& 7.63--\textbf{1.03}	& 20.4--\textbf{2.76} 	\\
			 	& f	& 0.68	& 0.23--\textbf{0.02}		& 0.50--\textbf{0.05}		& 7.12--\textbf{0.77}	& 19.0--\textbf{2.06} 	\\
			 	& g	& 1.34	& 0.14--\textbf{0.01}		& 0.32--\textbf{0.03}		& 7.62--\textbf{0.61}	& 20.3--\textbf{1.62} 	\\
				& h	& 0.41	& 0.02--\textbf{$<$0.01}	& 0.05--\textbf{$<$0.01}	& 2.82--\textbf{0.14}	& 7.52--\textbf{0.37}	\\
\hline
\end{tabular} 
\label{waterloss_photodisss}
\end{table*}

\subsection{Outgassing}   
 
Water is contained within a rocky planet's mantle in the form of hydrated minerals, as unbound fluids, or in melt. This water can be released to the surface through volcanic activity. Such outgassing processes are very different during the early stages of planet evolution ($<$ 10-100\,Myr, see \citealt{Solomatov2007}), where the surface and a significant fraction of the planet's mantle could be molten (magma ocean phase), and a later stage, where the largest fraction of the planet's mantle is solid. In the following sections, we will discuss the amounts of outgassed water during the magma ocean phase of the TRAPPIST-1 planets, assuming they are rocky, and we compute the water outgassing rates for the later stage of subsolidus convection as a function of time. 

\subsubsection{Outgassing from a magma ocean}  

At the early stages of planet evolution, magma oceans can outgas large fractions of water. The amount of outgassed water can range from less than 1\% to 20\% of a terrestrial-like planet mass based on typical compositions (see \citealt{Elkins2008}). For the Earth (M$_\mathrm{Earth}$ = 5.972$\times$10$^{24}$\,kg), this would correspond to up to 800 oceans of water assuming chondritic CI meteorites as the planet's building blocks. This number is clearly an upper limit, considering the fact that the Earth (and possibly the planets in the TRAPPIST-1 system), might have formed from much drier (intermediate size) planetary bodies or lost significant amounts of water in the impact-driven formation processes. Furthermore, the model of \citet{hamano2013} suggests that planets with steam atmospheres at orbits with a stellar influx larger than about 300 W\,m$^{-2}$ (~the Earth's current incoming solar radiation) would have much longer magma ocean phases and could hence possibly outgas much more of their initial water at an early stage in the first few 10-100 Myr, and thus end up much drier than planets at greater distances from their host star. Based on the heat fluxes for the TRAPPIST-1 planets today (see \citealt{Gillon2017}), planets b-d would fall in orbits with such elevated heat fluxes. However, at this point in time, it is almost impossible to estimate the initial water inventory after the planet formation phase (including magma ocean and a possible later delivery of water), because of a lack of detailed understanding of what the early radiation environment of TRAPPIST-1 looked like, whether the planets migrated inwards from a greater distance during this epoch (at subcritical lower stellar flux levels), whether any of the planets had a steam atmosphere (hydrogen- and methane-rich atmospheres would have no outgoing radiation limit to slow cooling, see \citealt{elkins2013}; and hence not fall under the \citealt{hamano2013} dichotomy), and whether any water was delivered by impactors after the magma ocean phase. Therefore, what we can constrain at this point in time are the limits to the outgassed amount of water during the magma ocean phase, and the later maximum subsolidus outgassing of water – assuming plausible ranges of post-magma-ocean water content in the planets' mantles (see next sections).

\subsubsection{Outgassing after the magma ocean phase} 

After the magma ocean phase, rocky planets cool mainly through subsolidus convection and outgas water through volcanism. Volcanic activity and outgassing are driven by the formation and transport of melt within the planet's subsurface, which depend on the thermal profile, the melting temperature of rock, and on the advective transport of rock parcels within a planet's lithosphere. 

\subsubsection{Methods}
\label{sec:meth_outgas}

Melt is formed only where the temperature exceeds the melting temperature of rock. We, therefore, have to model 1) the interior depth-dependent thermal evolution of a planet and know 2) the melting curve of mantle rock in order to assess a planet's ability to generate melt.
\begin{enumerate}
\item The thermal histories of planets are computed with an extended 1D boundary layer model (\citealt{Stamenkovic2012}), which agrees well with spherical 2D thermal evolution models (i.e., \citealt{Huttig2008}). The thermal evolution is described by two thermal boundary layers, which drive thermal convection and are used to parameterize the heat flux out of the core and out of the convective mantle, which is fed by secular cooling and radiogenic heat sources decaying in time. For simplicity, we assume the Earth's radiogenic heat content to obtain a first insight into the TRAPPIST-1 planets; we do not expect our conclusions to significantly vary for alternative values based on some first tests.\\
We try to mimic an initially fully molten mantle by fitting the initial upper mantle temperature T$_\mathrm{m}$(0) to the solidus temperature of peridotite, T$_\mathrm{melt,peridotite}$ (see below). For the highest initial core-mantle boundary (CMB) temperature, T$_\mathrm{c}$(0), we use the melting temperatures of MgSiO$_\mathrm{3}$ perovskite from \citet{Stamenkovic2011}. This assumption allows us to fluently connect to an initial early magma ocean stage. All scaling relations and parameters can be found in \citet{Stamenkovic2012}.
\item Our melt model has been described in detail in \citet{Stamenkovic2014}, and for the melting temperature in the upper mantle (where the melt that drives outgassing is produced) our model uses the solidus for Earth-like peridotite, T$_\mathrm{melt,peridotite}$. T$_\mathrm{melt,peridotite}$ is obtained by fitting the data from \citet{Herzberg2000}, \citet{Zerr1998}, and \citet{Fiquet2010} (see Eq. 7 in \citealt{Stamenkovic2014}). \\
Water can have a significant effect on reducing the solidus temperature of mantle rock (e.g., \citealt{Asimow2004}; \citealt{Aubaud2004}; \citealt{Grove2009}; \citealt{Hirschmann2006}). \citet{Asimow2004} compute melting curves for water-under-saturated and water-saturated peridotite. Their water-saturated melting curve is close to a prediction based on the homologous temperature approach (for details, see, e.g., \citealt{Katayama2008}; \citealt{Stamenkovic2011}). In the homologous temperature model, the melting temperature change corresponds to the enthalpy change of diffusion creep, E$^{*}$, (at smaller pressures this approximately corresponds to the activation energy change), so that T$_\mathrm{melt,dry}/$T$_\mathrm{melt,wet}$ = E$^{*}_\mathrm{dry}/$E$^{*}_\mathrm{wet}$. This corresponds to a melting point reduction of $\sim$20\% due to water saturation when we use the model of \citet{Karato1993} for activation energies of dry and wet olivine. We, therefore, use our water-saturated homologous temperature based melting curve as the reference melting curve for water-saturated upper mantle rock.
\end{enumerate}

The upper mantle water concentrations are thought to generally be below water saturation levels, typically between 50-200\,ppm (partially up to 1000\,ppm, still not saturated) (\citealt{Aubaud2004}; \citealt{Hirschmann2006}), at subduction zones locally over-saturated (e.g., \citealt{Grove2009}), and plumes are found to contain about 300-1000\,ppm of water (\citealt{Hirschmann2006} for review). The storage capacity of olivine on the other hand has been estimated to increase with depth from $\sim$25\,ppm at 10\,km to $\sim$1300\,ppm at 410\,km for the Earth, strongly varying with water fugacity and hence temperature and depth. It is, however, possible that this value is about $\sim$3-3.5 times too small, leading to more than $\sim$0.4 weight \% of water for mantle rock (see \citealt{Hirschmann2006} and references therein). On the other hand, based on geochemical constraints on K$_{2}$O/H$_{2}$O ratios in basalts (\citealt{Hirschmann2006} for review), the bulk water content is estimated to be between 500-1900\,ppm. For our first order of magnitude estimate for the TRAPPIST-1 system, we assume no depth-dependence of water content or storage capability and use average bulk values of 500\,ppm (minimal value bulk mantle) to 0.4\% (upper saturation value) for mantle rock. The rheology is fixed to a Newtonian-type viscosity for a wet bulk mantle based on \citet{Karato1993}. \\

We vary the pressure dependence (activation volume V$^{*}$) of the mantle viscosity, from V$^{*}$=0 to values calculated in \citet{Stamenkovic2011}. We propagate this uncertainty in mantle viscosity throughout all calculations. Furthermore, we follow the probabilistic approach of \citet{Stamenkovic2016}, where we also propagate an uncertainty in our heat flux (Nusselt) scaling parameter $\beta$, allowing it to vary between 0.2-1$/$3. By accounting for as many uncertainties as possible, we make sure that our results are as robust as possible. Within this parameter space, we highlight a favored model with a value of $\beta\sim$0.3 for the Nusselt-Rayleigh parameter and a pressure-dependent activation volume as computed in \citet{Stamenkovic2011} – suggesting that this standard model best represents the thermal evolution of rocky planets of variable core size between 0.1-2 Earth masses, representative of the possible refractory planet masses for the TRAPPIST-1 planets. We fix the surface temperature to 298\,K, as surface temperature variations found in the TRAPPIST-1 system today have no significant impact on our results (unless surface temperatures are above ~500-700 K).\\

Knowing a planet's ability to generate melt at depth and in time is however not sufficient to calculate whether that parcel of melt can be brought to the surface leading to potential outgassing. The latter depends strongly on two factors, 1) the density cross-over pressure of mantle rock and 2) the tectonic mode of a planet. 
\begin{enumerate}
\item Melt generated at depth will rise to the surface as long as the density of melt is smaller than that of the surrounding solid rock. However, on the Earth typical mantle rock at pressures above 12\,GPa does not rise to the surface due to the density-cross over, where melt becomes denser than surrounding solid rock (\citealt{Ohtani1995}). We note that this pressure value of 12\,GPa varies with rock composition and especially water content (\citealt{Jing2009}). Hence, we use the terrestrial value only as a reference point to explore whether the density cross-over pressure might affect outgassing on the TRAPPIST-1 planets. Moreover, we do not account for any other mechanisms that could cause intrusive volcanism.
\item The tectonic mode has two end members: plate tectonics (PT), as found on the Earth, and stagnant lid (SL) convection, as found on modern-day Mars. In the following, we will model outgassing for stagnant lid planets. Modeling outgassing in the plate tectonics mode is too sensitive to planet properties that we do not yet know from the TRAPPIST-1 planets, and hence we leave this to future work and refer to \citet{Schaefer2015} for a more detailed discussion on outgassing on plate tectonics worlds.
\end{enumerate}

For the TRAPPIST-1 planets, the uncertainties in mass are yet too large to infer much structural or compositional detail. At this point in time, what we can do is model the thermal evolution, melt generation, and water outgassing assuming a terrestrial (refractory) planetary body with a mass between 0.1-2 Earth masses with variable iron core sizes from 0-65\% (corresponding to core-less to Mercury-structured) in the stagnant lid mode - and put the TRAPPIST-1 planets in context with these results. Also, we note that we do not include tidal heating at this point in time. To make significant conclusions about the effects of tidal heating on the thermal evolution of the TRAPPIST-1 planets, we need much better constraints on planet masses, their volatile content, and their structures. Therefore, better constraints on planet masses will significantly improve our predictions in the near future.

\subsubsection{Results}

\begin{figure}[htbp!]
        \centering
        \includegraphics[width=\linewidth]{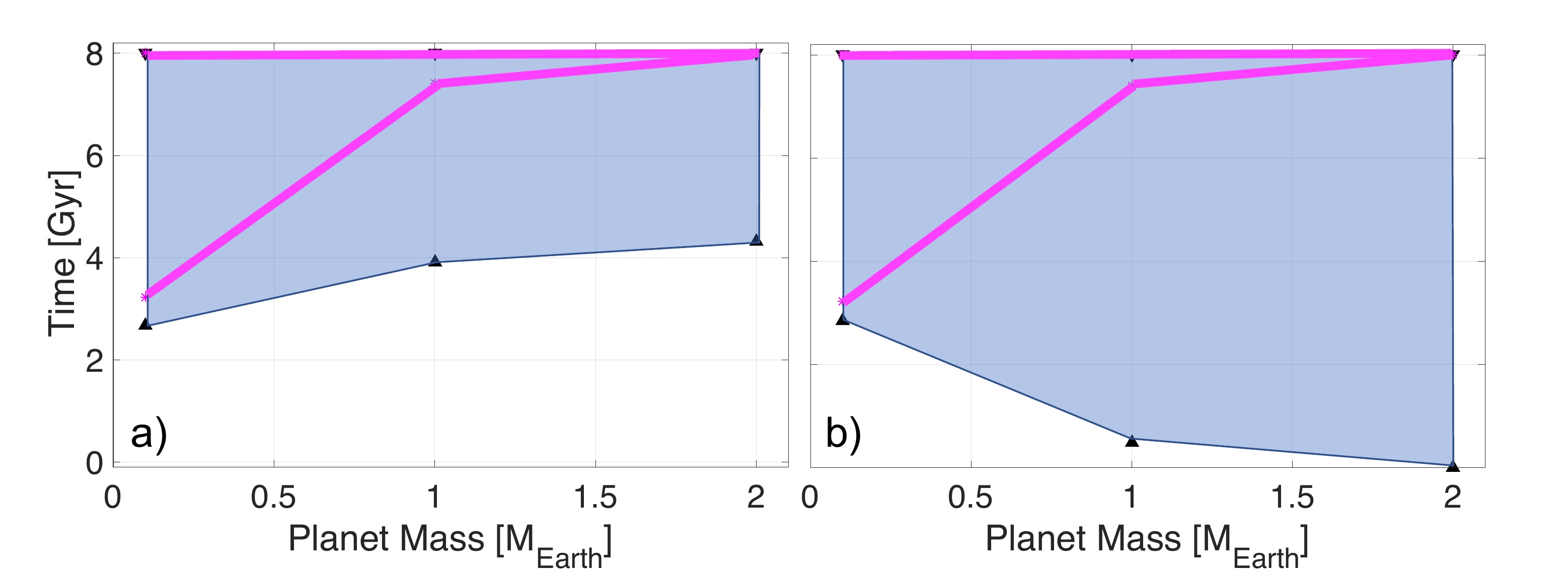}
        \caption{Duration of outgassing: As a function of refractory planet mass, we show the minimal duration of outgassing without (a) and with (b) consideration of an Earth-like reference density cross-over, including all uncertainties specified in the methods (Sect.~\ref{sec:meth_outgas}) in shaded blue. This uncertainty range is reduced to the domain in between the pink lines when considering only our standard model. When the density cross-over is considered, more massive planets can lack any extrusive volcanism due to the melt source region being too deep (and hence at too high pressures). However, for our standard model, we find that for all cases, more massive planets can outgas longer.}
        \label{dur_outgas}
        \end{figure}

\begin{figure}[htbp!]
        \centering
        \includegraphics[width=\linewidth]{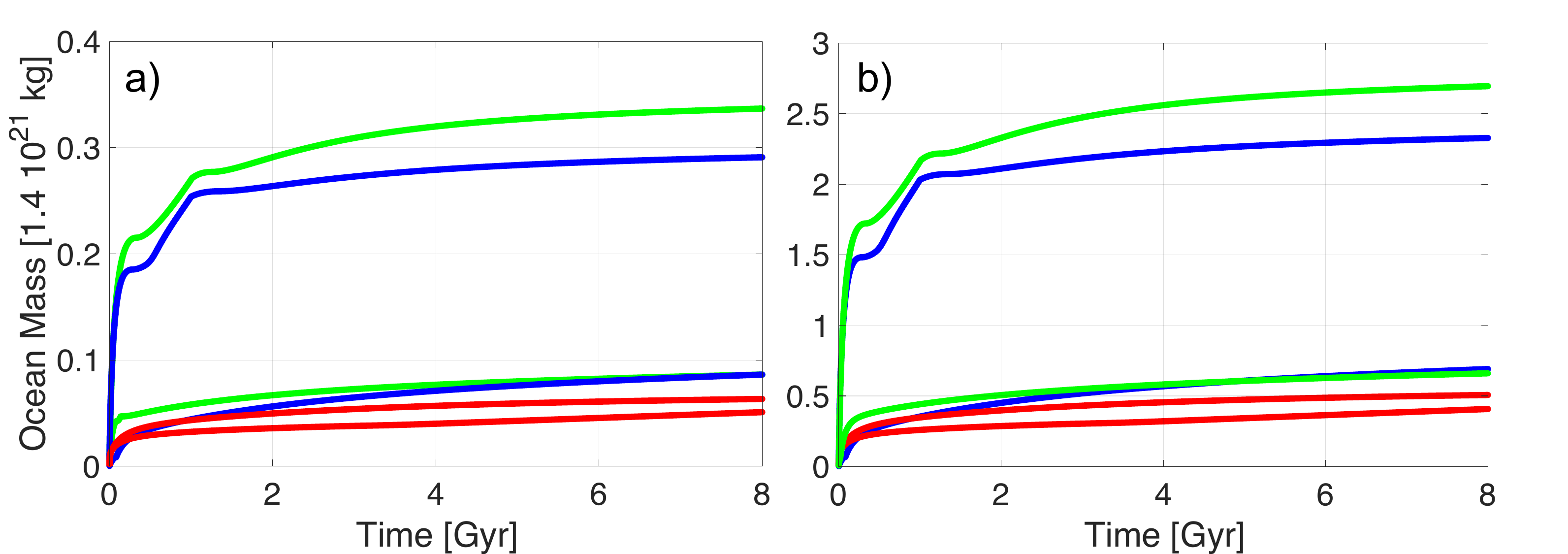}
        \caption{Outgassed water: We plot the range for the amount of outgassed water as a function of planet age for planets of refractory planet mass M=0.1 (red), 1 (blue), and 2 (green). The two colored lines delimit in each case the range of uncertainties in planet structure (from core-less to Mercury-structured) for our standard model. Figure (a) assumes 500\,ppm of water in the planet's mantle and (b) saturation levels of 0.4 weight \%. }
        \label{oceans_outgas}
        \end{figure}

We find that after the magma ocean phase, the TRAPPIST-1 planets can outgas significant amounts of water, especially the more massive ones. We show in Fig.~\ref{dur_outgas} - whilst accounting for significant uncertainties in structure and model parameters - the range of minimal planet ages where outgassing can occur as a function of planet mass assuming that the system is not older than 8\,Gyr. Within this domain, we also plot the solution for our standard model (in pink) without and with consideration of the density cross-over at 12\,GPa. We see in Fig.~\ref{dur_outgas} that the minimal ages during which outgassing can occur vary largely (mainly modulated by core fraction and Nusselt-Rayleigh parameter uncertainty). However, our standard model shows a robust behavior, independent of density crossover, suggesting that planets formed from more massive refractory parent bodies will be able to outgas much longer. Fig.~\ref{oceans_outgas} (showing the outgassed amount of water, in Earth oceans, for 500\,ppm and 0.4 weight \% of water respectively) also exemplifies that planets with more massive refractory parent bodies can outgas more water and outgas that water at much later times in their evolution.
	
	Combining this finding with Fig.~\ref{trappist_HZin_00875_Msun_all7planets}, which shows that planets within the orbits of TRAPPIST-1d and TRAPPIST-1h enter the HZ within 100\,Myr to a few hundred Myr, and considering that the largest atmospheric loss processes occur before entering the HZ, suggests that especially planets farther away from TRAPPIST-1 and planets that are more massive could deliver up to 1-2 ocean masses of water after they entered the HZ. This emphasizes that late-stage geophysical outgassing might be a critical component helping to sustain habitable environments within the TRAPPIST-1 system.

\section{Discussion}
\label{sec:conclu}

We observed the Ly-$\alpha$ line of TRAPPIST-1 with HST/STIS in December 2016, at the time of planet c transiting. When compared with previous observations obtained in September and November 2016, this new measurement revealed that the stellar line evolved significantly in the last visit. It shows an increased emission with broader wings, which might trace an increase in the temperature of the stellar chromosphere. The relation from \citet{Youngblood2016} between the stellar rotation period and  Ly-$\alpha$ surface flux of early-type M dwarfs predicts significantly stronger Ly-$\alpha$ emission than we detected for TRAPPIST-1. Our measured Ly-$\alpha$ fluxes would correspond to rotation periods of $\sim$84 days (based on Visits 1-3 Ly-$\alpha$ flux) and $\sim$54 days (based on Visits 4 Ly-$\alpha$ flux), whereas the rotation period from from K2 photometric data is $\sim$3.3\,days (\citealt{Luger2017}). This is in contrast to our detection of N$_\mathrm{v}$ emission, which we find to be consistent with the previously measured X-ray flux. Together these observations support our hypothesis that TRAPPIST-1 has a weak chromosphere compared to its transition region and corona (\citealt{Wheatley2016}, B17).\\

The spectra in December 2016 are subjected to a strong airglow contamination, which could have biased the extraction of TRAPPIST-1 Lyman-$\alpha$ line. Nonetheless, a careful analysis of the stellar line shape tentatively suggests an absorption from neutral hydrogen at high velocity in the blue wing. This signature does not seem to correlate with the transit of TRAPPIST-1c, but could originate from a system-wide neutral hydrogen cloud sustained by the evaporation of several planets, and shaped by the very low radiation pressure and photoionization from TRAPPIST-1. Alternatively, the peculiar shape of TRAPPIST-1 Ly-$\alpha$ line in this epoch could result from physical mechanisms specific or magnified in ultracool dwarfs. In any case the long-term, and possibly short-term variability in the intrinsic Ly-$\alpha$ line of TRAPPIST-1 prevents us from constraining the presence of a putative hydrogen exosphere around planet c, and calls for an ongoing monitoring of the star both outside and during all planets transits.\\

Combining all measurements of TRAPPIST-1 Ly-$\alpha$ and X-ray emissions, we estimated the present day XUV irradiation of the planets. Using simple assumptions on the evolution of the irradiation over time, we calculated the history of hydrodynamic water loss from the planets in the energy-limited regime. With our current knowledge of the TRAPPIST-1 system, the major uncertainty on the water loss estimates comes from the uncertainty on the planet masses, rather than on the XUV luminosity. Setting the masses to their nominal estimates from \citet{Gillon2017} we found that planets g and closer-in could have lost more than 20 Earth oceans through hydrodynamic escape, if the system is as old as 8\,Gyr. Planets b, c, and possibly d could still be in a runaway phase, but if water loss drops down significantly within and beyond the HZ, planets e, f, g, and h might have lost less than 3 Earth oceans. We caution that our water loss estimates were derived in a simplified framework, and should be considered as upper limits because our assumptions likely maximize the XUV-driven escape. We refer the reader to the section 6 of \citealt{Bolmont2017} for more details about these limitations. Furthermore, we found that photodissociation of water in the upper atmospheres of the TRAPPIST-1 planets is likely to be the limiting process, as hydrogen is produced at a lower rate than it is lost through hydrodynamic escape rate. Photolysis efficiency is expected to be lower than about 20\%, in which case all planets but TRAPPIST-1b and c could still harbor significant amounts of water. Naturally this also depends on the age of the system and whether the planets formed with a small water content (as suggested by planetary formation models, \citealt{Ormel2017}) or as planet oceans (as hinted by the low densities of the outer planets, especially TRAPPIST-1f; \citealt{Gillon2017}, \citealt{Wang2017}). We have also shown that late-stage outgassing could contribute significant amounts of water after the planets have entered the HZ. The amount of water outgassed after a few hundred Myr (after which TRAPPIST-1h to e, and possibly d, have entered the HZ), is greater for planets with more massive refractory parent bodies. Improving our constraints on the planet masses of the TRAPPIST-1 system will therefore significantly improve our understanding of the variable outgassing capabilities of the TRAPPIST-1 planets and hence the current state of their atmospheres. \\

Our study focused on hydrodynamic water loss driven by the quiescent stellar XUV irradiation. We also investigated the effects of photolysis and outgassing, but we did not account for other physical mechanisms competing between the erosion of the atmospheres and their replenishing. For example the presence of other gases in the atmosphere would act to slow down the hydrodynamic outflow (just as in our model oxygen atoms can exert a drag on the hydrogen flow), although the background atmosphere could also be exposed to evaporation. We refer the reader to section 5 of \citealt{Ribas2016} for an extensive discussion about the loss of the background atmosphere and other escape processes. Stellar flares could occasionally increase the energy input into the atmospheres, enhancing the escape rate. However, no flaring activity was observed in TRAPPIST-1 X-ray and FUV observations, and the flares detected in optical (\citealt{Vida2017}, \citealt{Luger2017}) and infrared (\citealt{Gillon2016}, \citealt{Gillon2017}) wavelengths point toward a low activity, with weak flares once every few days and stronger flares once every two to three months. The planetary atmospheres might also have been eroded by the stellar wind of TRAPPIST-1, especially when it was more active during the early phases of the system. Assuming that the planets are unmagnetized, \citealt{Dong2017} derived upper limits on the atmospheric ion escape driven by the stellar wind. Tidal heating (due to the planets proximity to the star and the mutual dynamical interactions maintaining their slightly eccentric orbits) could maintain significant magnetospheres, able to protect the planets from a putative stellar wind, or it could  suppress dynamo activity. Furtheremore, planetary magnetic fields can, depending on the specific interaction between planet and star, enhance or reduce atmospheric loss rates (e.g., \citealt{Strangeway2005}). Currently we have no constraints on the ability of the TRAPPIST-1 planets to generate magnetic fields, as this significantly depends on planet composition and structure. We also lack knowledge about the winds of ultracool dwarfs and their evolution with the stellar magnetic field over time, although TRAPPIST-1 is so cold that its amosphere likely has a low level of ionization, resulting in a lower emission of charged particles than for a hotter star like Proxima Centauri (e.g. \citealt{Mohanty2002}). \\

Understanding the nature of the TRAPPIST-1 planets and their potential habitability will thus require the combination of theoretical studies (to better understand the time-dependent processes leading to geophysically-driven water outgassing, the atmospheric loss processes, and the role of magnetic fields in affecting them) with further photometric observations (to refine the planet radii and more importantly their masses through TTV), X-ray and UV observations of the stellar spectrum (to monitor the activity of the star and measure the high-energy irradiation of the planetary atmospheres, currently poorly known; \citealt{OmalleyJames2017}), transit spectroscopy in the FUV (to detect escaping hydrogen and possibly oxygen) and in the IR (to search for the signature of water in the bottom atmospheric layers).\\


\section*{Acknowledgements}
We give our thanks to the referee for a detailed review and useful suggestions. We thank A. Lecavelier Des Etangs for insightful discussion about habitability and extraction of the STIS spectra. We thank  S. Raymond, F. Selsis, and H. Wakeford for helpful comments about this study. This work is based on observations made with the NASA/ESA Hubble Space Telescope, obtained at the Space Telescope Science Institute, which is operated by the Association of Universities for Research in Astronomy, Inc., under NASA contract NAS 5-26555. These observations are associated with program HST-GO-14900, support for which was provided by NASA through a grant from the Space Telescope Science Institute. This work has been carried out partly in the frame of the National Centre for Competence in Research ``PlanetS'' supported by the Swiss National Science Foundation (SNSF). This project has received funding from the European Research Council (ERC) under the European Union's Horizon 2020 research and innovation programme (project FOUR ACES; grant agreement No 724427). V.B. acknowledges the financial support of the SNSF. V.S. has received support from the Simons Collaboration on the Origins of Life (338555, VS). PW is supported by STFC consolidated grant ST/P000495/1. M. Gillon, E. Jehin, and V. Van Grootel are Belgian F.R.S.-FNRS Research Associates. This work was partially supported by a grant from the Simons Foundation (PI Queloz, grant number 327127). L. D. acknowledges support from the Gruber Foundation Fellowship.
\bibliographystyle{aa} 
\bibliography{biblio} 

\end{document}